\documentclass[useAMS,usenatbib,usegraphicx]{mn2e}
\usepackage{fixltx2e}
\usepackage{lscape}

\def\aap{A\& A}
\def\aaps{A\& AS.}

\def\aj{{AJ}}
\def\apj{ApJ}
\def\apjl{ApJL}
\def\apjs{ApJS}
\def\apss{Ap\&SS}
\def\araa{ARA\&A}

\def\mnras{MNRAS}

\def\pasp{{PASP}}

\def\ala#1{$^{#1}$}

\newcommand{\barL}{\mbox{$\bar{L}$}}

\newcommand{\barf}{\mbox{$\bar{f}$}}
\newcommand{\barm}{\mbox{$\bar{m}$}}
\newcommand{\barM}{\mbox{$\bar{M}$}}

\newcommand{\samename}{\vrule height0.4pt depth0.0pt width1.0in \thinspace.}

 \title[Mass-loss tracers] {Tracers of stellar mass-loss. I. Optical
and near-IR colours and surface brightness fluctuations}

 \author[Gonz\'alez-L\'opezlira et al.]
{ \parbox{7.0in}{
Rosa A.\ Gonz\'alez-L\'opezlira\ala 1
  \thanks{e-mail: {\tt r.gonzalez@crya.unam.mx}},
Gustavo Bruzual A.\ala 2,
St\'ephane Charlot\ala {3,4}
Javier Ballesteros-Paredes\ala 1
and Laurent Loinard\ala 1 \\
} \\ \ala 1 Centro de Radioastronom\'ia y Astrof\'isica,
            Universidad Nacional Aut\'onoma de M\'exico, \\
            Apdo. Postal 72-3 (Xangari), Morelia,
            Michoc\'an 58089, M\'exico \\
     \ala 2 Centro de Investigaciones de Astronom\'{\i}a, Apartado Postal 264,
M\'erida 5101-A, Venezuela\\
\ala 3 UPMC Univ Paris 06, UMR7095, Institut d'Astrophysique de Paris,
F-75014, Paris, France\\
\ala 4 CNRS, UMR7095, Institut d'Astrophysique de Paris, F-75014, Paris,
France\\
}

\begin{document}
\date{Submitted to MNRAS, \today}
\pagerange{\pageref{firstpage}--\pageref{lastpage}} \pubyear{2009}

\maketitle

\label{firstpage}

\begin{abstract}

We present optical and IR integrated colours and SBF magnitudes, 
computed from stellar population synthesis models that include
emission from the dusty envelopes surrounding TP-AGB
stars undergoing mass-loss.  We explore the effects of varying the mass-loss rate by one
order of magnitude 
around the fiducial value, modifying accordingly both the stellar
parameters and the output spectra of the TP-AGB stars plus their dusty
envelopes.  The models are single burst, and range in age from a few
Myr to 14 Gyr, and in metallicity between $Z$ = 0.0001 and $Z$ = 0.07;
they combine new
calculations for the evolution of stars in the TP-AGB phase,
with star plus envelope SEDs produced with the
radiative transfer code DUSTY. 
We compare these models
to optical and near-IR data of single AGB stars and Magellanic star
clusters.  This comparison validates the current understanding of the
role of mass-loss in determining stellar parameters and spectra in the
TP-AGB. However,
neither broad-band colours nor SBF measurements in the optical or the
near-IR can discern global changes in the mass-loss rate of a stellar
population.  We predict that mid-IR SBF measurements can pick
out such changes, and actually resolve whether a relation between
metallicity and mass-loss exists.

\end{abstract}


Online-only material: machine-readable and VO tables

\section{Introduction.}

Stellar mass-loss is inseparable from stellar evolution and death.  It
is fundamental to inject enriched material into the interstellar
medium, and hence a major driver of the chemical enrichment and
evolution of galaxies. Whereas the energetic photons emitted by
massive, young stars destroy molecular gas, the dense outflows of
evolved stars return to the ISM both molecular gas and dust grains
where more H$_2$ can then form.  Mass-loss is especially important at
the tip of the asymptotic giant branch (AGB) for intermediate-mass
stars ($1 M_\odot \la M_{ZAMS} \la 9 M_\odot$).  Consequently,
mass-loss plays a central role also in shaping the AGB and planetary
nebula luminosity functions. It determines the white dwarf mass
spectrum and cooling times, and the minimum supernova (SN) progenitor
mass. Mass-loss thus influences the rate of SNe types II, Ib, and Ic, and possibly
determines the mass of SNe Type Ia progenitors and impacts their
frequency \citep{bowe91,will00}.

Thermally-pulsing AGB (TP-AGB) stars are the most luminous red stars
in intermediate-age stellar populations, and accordingly affect their
integrated properties.  The relevance of AGB stars contrasts with our
inadequate understanding of this evolutionary phase. The TP-AGB has
proven especially hard to model, on account of the thermal pulses they
suffer; the convective dredge-up of processed heavy elements to the
stellar surface; and the ejection of the stellar envelope that ends
the phase \citep{bruz03}.  There is an extensive bibliography dealing
with the difficulties in treating the TP-AGB
\citep[e.g.,][]{renz81,iben83,mari96,mara98,mara05}.  In particular,
the relation between the fundamental parameters (luminosity, $L$;
mass, $M$; and metallicity, $Z$) of the central star and the mass-loss
rate, $\dot M$, is not quite well understood and, therefore,
controversial.  Of course, it does not help that the dust in the
ejected envelopes precludes the direct observation of the stellar
photospheres.  Data with relatively good spatial resolution of the
dusty cocoons themselves, on the other hand, have been available only
in the last decade, with the arrival of the Infrared Space Observatory
(ISO) and, presently, the Spitzer Space Telescope.
 
Customarily, $\dot M$ has been described by the empirical Reimers' law
\citep{reim75,reim77}, written as $\dot M = \eta LR/M$; $R(L,M,Z)$ is
the stellar radius, and $\eta$ is a fitting parameter.  Modifications
to this law, motivated by the large scatter of observationally
determined mass-loss around the Reimers' relation, have been proposed
by, among others, \citet{baud83}, \citet{bloe95}, and \citet{groe94}.
Those working on the detailed modeling of mass-loss at the TP-AGB have
countered that, in actuality, stellar luminosity first increases at
constant mass until the star reaches a ``cliff" in the log $M$ vs.\
log $L$ plane; after this point, mass-loss depends much more steeply
on stellar parameters than stated by empirical relations, such that
the stellar envelope is lost nearly exponentially in time at roughly
constant luminosity (Bowen 1995; Willson 2000, her Fig.\ 7, and
references therein).  In this view, empirical relations are the result
of very strong selection effects, and only reflect the fact that
mass-loss rates are measurable for just a fraction of the stars
undergoing mass-loss at any given time.  Thus, in the log $\dot M$
vs.\ $LR/M$ plane, the so-called cliff and Reimers' relation are
almost coincident (see Fig.\ 8 of Willson 2000).  Before reaching the
cliff, stars have low, unmeasurable mass-loss rates; conversely, the
stage beyond the cliff is short-lived, and most likely stars will be
highly obscured behind a dusty envelope. A sample of stars with
observable mass-loss rates should include, according to
\citet{will00}, mainly objects within one dex in $\dot M$ of the
cliff.

The purpose of this paper is to produce 
population synthesis models that include different mass-loss rates in
the TP-AGB.  To this end, we combine stellar population evolutionary
models with theoretical spectral energy distributions (SEDs) that
take into account the radiative transfer in the dusty circumstellar
envelopes.  To begin with, the SEDs are calculated on the basis of the
mass-loss rates included in the evolutionary tracks; we then produce
SEDs for the same stellar types, but with mass-loss rates one order of
magnitude above and below the rates in the tracks, in order to explore
the whole range where mass-loss is observable in the optical. Finally,
we confront the resulting theoretical broad-band colours and
fluctuation magnitudes with optical and near-IR observations of AGB
stars and Magellanic star clusters.

\section{Stellar population synthesis models and $\dot M$.} \label{marigomodels}

In this paper we use a preliminary version of the Charlot \& Bruzual
(2009; CB09 henceforth) simple stellar population (SSP) evolutionary
synthesis models to compute isochrones in the age range from a few Myr
to 14 Gyr, and metal (helium) content $Z(Y)$ = 0.0001(0.26),
0.0004(0.26), 0.001(0.26), 0.002(0.26), 0.004(0.26), 0.008(0.26),
0.017(0.26), 0.04(0.30), and 0.07(0.34).  The CB09 models are formally
identical to the Bruzual \& Charlot (2003; BC03 hereafter) models, but
include several important improvements.  Firstly, 
up to $15 M_\odot$ and for the metal and He contents
indicated above, CB09 use the tracks from the models with updated input 
physics by \cite{bertl08}. For
stars more massive than $15 M_\odot$, in the range from $20$ to $120
M_\odot$, CB09 use the so-called Padova 1994 tracks
\citep{alon93,bres93,fa94a,fa94b,gira96}.

Secondly, in the CB09 models used in this paper, the TP-AGB evolution
of low- and intermediate-mass stars is followed according to the
prescription of \cite{mari07}.  This semi-empirical prescription
includes several important theoretical improvements over previous
calculations, and it has been calibrated using carbon star luminosity
functions in the Magellanic Clouds (MC) and TP-AGB lifetimes (star counts)
in MC clusters (we refer to the paper by Marigo \&
Girardi for details).  The reader should be aware that \cite{bertl08}
use a different set of TP-AGB tracks, also based on the \cite{mari07}
prescription, but extrapolated to different chemical compositions of
the stellar envelope. These sets of TP-AGB tracks are
un-calibrated, as pointed out by \cite{bertl08}, since no attempt was
made to reproduce the available observations.  CB09 will discuss the
differences introduced in the evolutionary models by the use of the
calibrated or the un-calibrated TP-AGB tracks from these authors.

Thus, the CB09 models discussed here use the tracks in the
\citet{bertl08} atlas up to the end of the AGB phase, and extend these
tracks with the results of \citet{mari07} to cover the TP-AGB phase.
\cite{gba07} has shown that models computed following the
\cite{mari07} prescription have brighter $K$-band magnitudes and
redder near-IR colours than other models, e.g., BC03, that use a
semi-empirical treatment of the TP-AGB evolution based on an older
empirical calibration of the lifetime of these stars, and an educated
guess of the mass associated to TP-AGB stars of a given luminosity.\footnote{We
refer to the papers by Marigo \& Girardi and BC03 for details. In particular,
\citet{gira07} derive constraints on TP-AGB lifetimes, while 
\citet{mari07} compare their theoretical initial-final mass relation
with empirical data of white dwarfs in open clusters and 
binary systems.} 

The Marigo \& Girardi prescription, as implemented by CB09, accounts
for 15 evolutionary stages in the TP-AGB (six in the O-rich phase, six
in the C-rich phase, and three in the superwind phase).  By contrast,
the BC03 models include only one evolutionary stage at each of these
phases.  The signature of TP-AGB stars, i.e., the red colour of the
integrated SSP around 1 Gyr, becomes more relevant in the CB09 models
than in previous computations by BC03 and other authors
\citep[e.g.,][]{cm06}. For $Z = 0.008$, the TP-AGB stars in the
CB09 models contribute close to a factor of two more light in the
$K$-band than in the BC03 models.  At maximum, the TP-AGB contributes
close to 70\% of the $K$-light in the CB09 model for $Z = 0.008$, but only 40\% in the
BC03 model. The peak $K$-band luminosity in the BC03 model occurs at around 1
Gyr, whereas in the CB09 model it stays high and close to constant
from 0.1 to 1 Gyr. 
The evolutionary rate is such that the total number of
TP-AGB stars present in the CB09 1 Gyr isochrone is 3.4 times larger
than the number of these stars in the BC03 models.  The TP-AGB stars
represent 0.012\% of the total number of stars in this population at
this age in the CB09 model, but only 0.0036\% in the BC03 model.  A
$10^6~M_\odot$ cluster contains 181 TP-AGB stars in the CB09 model,
but only 53 of these stars in the BC03 model.  
The TP-AGB stars are about 0.3 mag brighter at $K$ in the BC03
than in the CB09 isochrone. The net effect of all
these factors is an increase of roughly 90\% in the contribution of
TP-AGB stars to the total $K$ flux of the $Z=0.008$ SSP at this age.  See
\cite{gba07} for more details.

The CB09 isochrones provide, at any given age, the number of stars at
each of 310 positions in the $(L_{BOL}, T_{\rm eff})$ plane. To each
star along the isochrone, an SED is assigned from the \cite{west02}
stellar library for all stellar phases, except for the C-rich and 
superwind stages of the TP-AGB;
for these we use several options, including the DUSTY models already
mentioned above.  For each of the nine stellar metallicities 
considered, we have calculated CB09 models for four possible choices
of the SED assigned to TP-AGB stars in the superwind phase: (1) the
SEDs used in the BC03 and CB09 models (standard models hereafter),\footnote{
The spectra of TP-AGB stars have not been updated in the
preliminary version of the standard CB09 model used here with respect
to those used in BC03.  For C-type TP-AGB stars and stars in the
superwind phase at the end of the TP-AGB evolution, BC03 and CB09 use
period-averaged spectra based on models by \citet{schul99} and
observations of Galactic stars.  For O-rich TP-AGB stars and stars at
the tip of the RGB, BC03 and CB09 use the \citet{west02} atlas.  The
final version of CB09 will include new spectra for TP-AGB stars.} and
(2) model spectra computed with the code DUSTY for: $(a)$ dusty
envelopes that result from the fiducial $\dot M$ during these
evolutionary phases, $(b)$ dusty envelopes from $\dot M$ one order of
magnitude above fiducial, and $(c)$ dusty envelopes from $\dot M$ one
order of magnitude below fiducial.  The input spectra for the DUSTY
code are the same spectra used at these phases by CB09, albeit modified
according to the updated stellar parameters if the mass-loss-rate is
$\dot M \times 10$ or $\dot M / 10$. In all cases
we use the \citet{chab03} IMF.

We then compute the time evolution of several properties of
single-burst stellar populations, in particular their integrated
fluxes and fluctuation luminosities in the $V$, $B$, $R$, $I$, $J$,
$H$, and $K_s$ bands, as well as in the IRAC and MIPS wavelengths
observed by the Spitzer Space Telescope (Section~\ref{sec_theosbf}).\footnote {
Strictly speaking, the integrated fluxes and fluctuation luminosities of a stellar
population at a given time are not the product of the mass-loss rate
at that time (or at any time), but rather the effect of the total
mass-loss up to that point.}

Besides predicting stellar time evolution on the Hertzsprung-Russell
diagram, the new \citeauthor{mari07} models self-consistently estimate
pulsation modes and periods, changes in the chemical composition of
the stellar envelopes and, most relevant for the present work,
mass-loss rates owing to the pulsating, dust driven winds of O- and
C-rich AGB stars.  In the case of O-rich stars, \citet{mari07} derive
luminosities from $M$, $R$, and pulsation period ($P$); next, they
calculate stellar mass-loss rates according to the stars' evolutionary
slopes d (log $M$)/ d (log $L$), on the basis of the \citet{bowe91}
dynamical atmospheres for Miras including dust.\footnote{ d (log $M$)/ d
(log $L$) = 1, and $\dot M$ = 5.67 $\times 10^{-7} M_\odot$ yr$^{-1}$
along the ``cliff" line; this rate marks the onset of the ``superwind"
phase.}  The \citet{bowe91} models anticipate that mass-loss rates of
O-rich stars diminish with decreasing metallicity, due to a less
efficient dust production and a smaller photospheric radius at a fixed
luminosity.  To determine mass-loss rates of C-rich stars,
luminosities are calculated first from stellar temperature ($T_{\rm
eff}$), $M$, and $P$.  At low luminosities, mass-loss is driven mostly
by stellar pulsation, whereas radiation pressure on dust grains plays
a secondary role. However, at a critical luminosity that depends on
stellar mass, dust-driven superwinds take place.  For the superwind
phase, \citeauthor{mari07} calculate $\dot M (T_{\rm eff}, L, M, P,
C/O) $ based on the pulsating wind models by \citet{wach02}, although
including in addition an explicit dependence on the C/O ratio.  The
mass-loss at each evolutionary stage is then taken into account to
consistently determine the stellar parameters in subsequent phases.

Examples of predicted mass-loss rates are presented by \citet{mari07}
in their Figures 15 and 16. They also note that, for all cases and C/O
values: $(a)$ $\dot M$ is ultimately controlled by the changes in $L$
and $T_{\rm eff}$ linked to He-shell flashes, $(b)$ most of the
stellar mass-loss occurs during the high-luminosity but quiescent
stages that preceed thermal pulses, and $(c)$ the superwind regime is
achieved during fundamental mode pulsation.

To date, there are few cases where evolutionary synthesis models
include the effects of mass-loss in stellar spectra.  Among these,
\citet{lanc02}, \citet{mouh03}, and \citet{mara05} use averaged
observed SEDs, whereas \citet{bres98} and \citet{piov03} have
pioneered the use of analytical relations for the mass-loss rates and
wind terminal velocities that then allow them to model the spectra of
dusty circumstellar envelopes.  We have proceeded in a manner similar
to these latter works. For each stellar type in the TP-AGB, SEDs of
star plus envelope were produced with the radiative transfer code
DUSTY; as was mentioned before, SEDs were also processed for the same
stellar types, but with mass-loss rates one order of magnitude above
and below the rates in the tracks, in order to explore the whole range
where mass-loss is observable in the optical, according to
\citet{will00}.  A change in $\dot M$, in turn, entails variations in
$L$ and hence stellar lifetime, $R$, envelope and core masses,
$T_{\rm eff}$, pulsation period, dust-to-gas ratio, dust composition, and
C/O ratio.  A major challenge for this work is that we are aiming at
extrapolating mass-loss rates at a large range of metallicities, but
in fact a good calibration of all the parameters involved in
individual TP-AGB stars exists only for Galactic and Magellanic TP-AGB
stars.  Two possible routes are open: we can either limit our work to
these metallicities, or venture to make predictions for lower and
higher metallicities, with the clear caveat that most relations have
not been tested in these conditions.  We choose the second, and strive
to keep as close as possible to the procedures used by \citet{mari07}
to produce the original stellar tracks.

We present our results in the subsequent sections, while the
interested reader can follow our calculations in detail in the
Appendix. CB09 models (i.e., with fiducial mass-loss) are reported 
in tables 1 through 4. Optical and near-IR
colours as a function of age, for different chemical compositions, 
are presented in tables 1 and 2,
respectively without and with dusty envelopes; 
fluctuation amplitudes are listed in tables 3 and 4.

\begin{table*}
 \centering
 \begin{minipage}{280mm}
  \caption{Colours for CB09 models with fiducial mass-loss but without dusty envelopes}
  \begin{scriptsize}
  \begin{tabular}{@{}rrrrrrrr@{}}
\hline
 Age (Gyr)  & $B-V$  & $V-R$  & $V-I$  & $V-J$  & $V-H$  & $V-K_s$  & $V-K$ \\
\hline
\multicolumn {8}{|c|}{$Z = 0.017 \ \ Y = 0.26$}\\
\hline
     0.005  &    -0.142  &    -0.027  &    -0.043  &    -0.115  &    -0.031  &    -0.010  &    -0.017  \\
     0.006  &    -0.063  &     0.062  &     0.144  &     0.339  &     0.622  &     0.699  &     0.681  \\
     0.007  &    -0.059  &     0.070  &     0.175  &     0.443  &     0.793  &     0.900  &     0.878  \\
     0.008  &    -0.075  &     0.062  &     0.184  &     0.524  &     0.950  &     1.097  &     1.071  \\
     0.009  &    -0.133  &     0.020  &     0.161  &     0.604  &     1.146  &     1.354  &     1.324  \\
     0.010  &    -0.138  &     0.029  &     0.201  &     0.721  &     1.311  &     1.531  &     1.499  \\
     0.020  &    -0.071  &     0.096  &     0.348  &     0.967  &     1.606  &     1.816  &     1.781  \\
     0.030  &    -0.021  &     0.127  &     0.369  &     0.931  &     1.528  &     1.709  &     1.677  \\
     0.040  &     0.002  &     0.126  &     0.335  &     0.815  &     1.353  &     1.510  &     1.480  \\
     0.050  &     0.035  &     0.138  &     0.339  &     0.790  &     1.296  &     1.437  &     1.409  \\
     0.060  &     0.051  &     0.143  &     0.340  &     0.777  &     1.264  &     1.399  &     1.372  \\
     0.070  &     0.087  &     0.165  &     0.369  &     0.818  &     1.291  &     1.425  &     1.401  \\
     0.080  &     0.097  &     0.172  &     0.381  &     0.834  &     1.304  &     1.437  &     1.414  \\
     0.090  &     0.117  &     0.186  &     0.405  &     0.869  &     1.338  &     1.468  &     1.446  \\
     0.100  &     0.130  &     0.196  &     0.421  &     0.896  &     1.368  &     1.498  &     1.475  \\
     0.200  &     0.141  &     0.203  &     0.498  &     1.402  &     2.078  &     2.345  &     2.314  \\
     0.300  &     0.171  &     0.214  &     0.508  &     1.302  &     2.004  &     2.301  &     2.304  \\
     0.400  &     0.224  &     0.234  &     0.543  &     1.306  &     1.986  &     2.269  &     2.273  \\
     0.500  &     0.275  &     0.255  &     0.587  &     1.368  &     2.017  &     2.262  &     2.248  \\
     0.600  &     0.311  &     0.274  &     0.621  &     1.412  &     2.067  &     2.315  &     2.302  \\
     0.700  &     0.387  &     0.309  &     0.683  &     1.477  &     2.142  &     2.392  &     2.385  \\
     0.800  &     0.431  &     0.331  &     0.716  &     1.500  &     2.166  &     2.415  &     2.412  \\
     0.900  &     0.477  &     0.355  &     0.755  &     1.542  &     2.209  &     2.454  &     2.451  \\
     1.000  &     0.521  &     0.378  &     0.790  &     1.587  &     2.251  &     2.486  &     2.479  \\
     1.500  &     0.723  &     0.543  &     1.178  &     2.339  &     3.150  &     3.432  &     3.415  \\
     2.000  &     0.707  &     0.499  &     1.010  &     2.035  &     2.762  &     3.006  &     2.981  \\
     3.000  &     0.781  &     0.537  &     1.066  &     2.100  &     2.831  &     3.069  &     3.045  \\
     4.000  &     0.832  &     0.561  &     1.099  &     2.142  &     2.876  &     3.109  &     3.085  \\
     5.000  &     0.854  &     0.571  &     1.113  &     2.159  &     2.894  &     3.125  &     3.102  \\
     6.000  &     0.874  &     0.581  &     1.124  &     2.171  &     2.907  &     3.135  &     3.112  \\
     7.000  &     0.885  &     0.586  &     1.129  &     2.170  &     2.903  &     3.127  &     3.103  \\
     8.000  &     0.901  &     0.594  &     1.142  &     2.194  &     2.930  &     3.153  &     3.128  \\
     9.000  &     0.907  &     0.597  &     1.145  &     2.195  &     2.928  &     3.149  &     3.123  \\
    10.000  &     0.922  &     0.604  &     1.159  &     2.217  &     2.953  &     3.174  &     3.148  \\
    11.000  &     0.929  &     0.607  &     1.163  &     2.219  &     2.955  &     3.175  &     3.149  \\
    12.000  &     0.938  &     0.611  &     1.170  &     2.227  &     2.963  &     3.183  &     3.156  \\
    13.000  &     0.949  &     0.617  &     1.179  &     2.240  &     2.977  &     3.197  &     3.171  \\
    13.500  &     0.955  &     0.619  &     1.183  &     2.247  &     2.985  &     3.204  &     3.178  \\
\hline
\end{tabular}
\end{scriptsize}
\end{minipage}
{(This table is available in its entirety in the online journal, and at CDS in machine-readable format.
Values for solar metallicity and helium content are shown here for guidance regarding its form and content.)}\\
\label{tabla1}
\end{table*}

\begin{table*}
 \centering
 \begin{minipage}{280mm}
  \caption{Colours for CB09 models with fiducial mass-loss plus dusty envelopes}
  \begin{scriptsize}
  \begin{tabular}{@{}rrrrrrrr@{}}
\hline
 Age (Gyr)  & $B-V$  & $V-R$  & $V-I$  & $V-J$  & $V-H$  & $V-K_s$  & $V-K$ \\
\hline
\multicolumn {8}{|c|}{$Z = 0.017 \ \ Y = 0.26$}\\
\hline
     0.005  &    -0.142  &    -0.027  &    -0.043  &    -0.115  &    -0.031  &    -0.010  &    -0.017  \\
     0.006  &    -0.063  &     0.062  &     0.144  &     0.339  &     0.622  &     0.699  &     0.681  \\
     0.007  &    -0.059  &     0.070  &     0.175  &     0.443  &     0.793  &     0.900  &     0.878  \\
     0.008  &    -0.075  &     0.062  &     0.184  &     0.524  &     0.950  &     1.097  &     1.071  \\
     0.009  &    -0.133  &     0.020  &     0.161  &     0.604  &     1.146  &     1.354  &     1.324  \\
     0.010  &    -0.138  &     0.029  &     0.201  &     0.721  &     1.311  &     1.531  &     1.499  \\
     0.020  &    -0.071  &     0.096  &     0.348  &     0.967  &     1.606  &     1.816  &     1.781  \\
     0.030  &    -0.021  &     0.127  &     0.369  &     0.931  &     1.528  &     1.709  &     1.677  \\
     0.040  &     0.002  &     0.126  &     0.335  &     0.815  &     1.353  &     1.510  &     1.480  \\
     0.050  &     0.035  &     0.138  &     0.339  &     0.790  &     1.296  &     1.437  &     1.409  \\
     0.060  &     0.051  &     0.143  &     0.340  &     0.777  &     1.264  &     1.399  &     1.372  \\
     0.070  &     0.087  &     0.165  &     0.369  &     0.818  &     1.291  &     1.425  &     1.401  \\
     0.080  &     0.097  &     0.172  &     0.381  &     0.834  &     1.304  &     1.437  &     1.414  \\
     0.090  &     0.117  &     0.186  &     0.405  &     0.869  &     1.338  &     1.468  &     1.446  \\
     0.100  &     0.130  &     0.196  &     0.421  &     0.896  &     1.368  &     1.498  &     1.475  \\
     0.200  &     0.143  &     0.206  &     0.480  &     1.494  &     2.324  &     2.581  &     2.609  \\
     0.300  &     0.174  &     0.215  &     0.495  &     1.325  &     2.106  &     2.395  &     2.425  \\
     0.400  &     0.228  &     0.233  &     0.527  &     1.282  &     2.021  &     2.276  &     2.297  \\
     0.500  &     0.279  &     0.252  &     0.557  &     1.290  &     1.983  &     2.172  &     2.179  \\
     0.600  &     0.315  &     0.269  &     0.586  &     1.325  &     2.021  &     2.213  &     2.220  \\
     0.700  &     0.392  &     0.301  &     0.638  &     1.369  &     2.062  &     2.261  &     2.270  \\
     0.800  &     0.435  &     0.320  &     0.664  &     1.378  &     2.063  &     2.265  &     2.277  \\
     0.900  &     0.482  &     0.342  &     0.698  &     1.408  &     2.088  &     2.287  &     2.298  \\
     1.000  &     0.526  &     0.364  &     0.731  &     1.451  &     2.122  &     2.305  &     2.310  \\
     1.500  &     0.746  &     0.520  &     1.077  &     2.151  &     2.999  &     3.192  &     3.188  \\
     2.000  &     0.711  &     0.504  &     0.987  &     2.081  &     2.872  &     3.054  &     3.045  \\
     3.000  &     0.783  &     0.537  &     1.052  &     2.144  &     2.926  &     3.111  &     3.094  \\
     4.000  &     0.834  &     0.561  &     1.095  &     2.209  &     2.998  &     3.181  &     3.159  \\
     5.000  &     0.856  &     0.571  &     1.110  &     2.215  &     2.997  &     3.186  &     3.163  \\
     6.000  &     0.874  &     0.581  &     1.120  &     2.207  &     2.977  &     3.172  &     3.148  \\
     7.000  &     0.886  &     0.586  &     1.124  &     2.193  &     2.951  &     3.147  &     3.123  \\
     8.000  &     0.902  &     0.594  &     1.137  &     2.211  &     2.967  &     3.164  &     3.140  \\
     9.000  &     0.908  &     0.597  &     1.140  &     2.206  &     2.956  &     3.154  &     3.129  \\
    10.000  &     0.922  &     0.604  &     1.155  &     2.230  &     2.983  &     3.182  &     3.157  \\
    11.000  &     0.929  &     0.607  &     1.160  &     2.233  &     2.985  &     3.186  &     3.160  \\
    12.000  &     0.938  &     0.611  &     1.167  &     2.242  &     2.994  &     3.195  &     3.169  \\
    13.000  &     0.950  &     0.617  &     1.176  &     2.255  &     3.008  &     3.210  &     3.184  \\
    13.500  &     0.955  &     0.619  &     1.181  &     2.261  &     3.014  &     3.217  &     3.191  \\
\hline
\end{tabular}
\end{scriptsize}
\end{minipage}
{(This table is available in its entirety in the online journal, and at CDS in machine-readable format.
Values for solar metallicity and helium content are shown here for guidance regarding its form and content.)}
\label{tabla2}
\end{table*}

\begin{table*}
 \centering
 \begin{minipage}{280mm}
  \caption{SBF amplitudes for CB09 models with fiducial mass-loss but without dusty envelopes}
  \begin{scriptsize}
  \begin{tabular}{@{}rrrrrrrrr@{}}
\hline
 Age (Gyr)  & $\barM_B$  & $\barM_V$  & $\barM_R$  & $\barM_I$  & $\barM_J$  & $\barM_H$  & $\barM_{K_s}$  & $\barM_K$ \\
\hline
\multicolumn {9}{|c|}{$Z = 0.017 \ \ Y = 0.26$}\\
\hline
     0.005  &    -7.637  &    -7.809  &    -8.061  &    -8.340  &    -8.999  &    -9.510  &    -9.644  &    -9.617  \\
     0.006  &    -7.446  &    -7.697  &    -8.168  &    -8.678  &    -9.705  &   -10.389  &   -10.559  &   -10.521  \\
     0.007  &    -7.171  &    -7.415  &    -7.887  &    -8.432  &    -9.546  &   -10.286  &   -10.487  &   -10.446  \\
     0.008  &    -6.885  &    -7.052  &    -7.457  &    -8.033  &    -9.295  &   -10.137  &   -10.390  &   -10.347  \\
     0.009  &    -6.427  &    -6.470  &    -6.701  &    -7.330  &    -8.979  &    -9.998  &   -10.330  &   -10.286  \\
     0.010  &    -6.096  &    -6.092  &    -6.352  &    -7.155  &    -8.963  &    -9.984  &   -10.313  &   -10.270  \\
     0.020  &    -4.936  &    -4.972  &    -5.263  &    -6.051  &    -7.692  &    -8.698  &    -8.999  &    -8.954  \\
     0.030  &    -4.390  &    -4.500  &    -4.818  &    -5.491  &    -6.951  &    -7.945  &    -8.232  &    -8.186  \\
     0.040  &    -3.884  &    -4.049  &    -4.370  &    -4.973  &    -6.338  &    -7.338  &    -7.621  &    -7.573  \\
     0.050  &    -3.382  &    -3.658  &    -4.031  &    -4.624  &    -5.937  &    -6.929  &    -7.205  &    -7.156  \\
     0.060  &    -2.990  &    -3.367  &    -3.795  &    -4.406  &    -5.737  &    -6.742  &    -7.027  &    -6.977  \\
     0.070  &    -2.251  &    -2.832  &    -3.385  &    -4.052  &    -5.581  &    -6.665  &    -7.028  &    -6.975  \\
     0.080  &    -1.895  &    -2.547  &    -3.172  &    -3.883  &    -5.510  &    -6.626  &    -7.017  &    -6.964  \\
     0.090  &    -1.629  &    -2.341  &    -3.019  &    -3.754  &    -5.397  &    -6.517  &    -6.914  &    -6.860  \\
     0.100  &    -1.326  &    -2.066  &    -2.799  &    -3.577  &    -5.291  &    -6.428  &    -6.837  &    -6.782  \\
     0.200  &    -0.291  &    -0.671  &    -1.485  &    -3.217  &    -6.837  &    -8.026  &    -8.527  &    -8.478  \\
     0.300  &     0.083  &    -0.254  &    -1.139  &    -2.915  &    -6.166  &    -7.531  &    -8.121  &    -8.157  \\
     0.400  &     0.508  &     0.117  &    -0.786  &    -2.666  &    -5.760  &    -7.132  &    -7.724  &    -7.766  \\
     0.500  &     0.812  &     0.339  &    -0.595  &    -2.622  &    -5.698  &    -6.981  &    -7.496  &    -7.492  \\
     0.600  &     0.992  &     0.435  &    -0.555  &    -2.597  &    -5.600  &    -6.887  &    -7.406  &    -7.410  \\
     0.700  &     1.364  &     0.619  &    -0.483  &    -2.525  &    -5.359  &    -6.664  &    -7.198  &    -7.221  \\
     0.800  &     1.552  &     0.694  &    -0.461  &    -2.473  &    -5.199  &    -6.523  &    -7.072  &    -7.109  \\
     0.900  &     1.721  &     0.744  &    -0.447  &    -2.419  &    -5.072  &    -6.405  &    -6.967  &    -7.012  \\
     1.000  &     1.890  &     0.805  &    -0.405  &    -2.337  &    -4.968  &    -6.300  &    -6.869  &    -6.912  \\
     1.500  &     1.985  &     0.421  &    -0.890  &    -2.793  &    -5.305  &    -6.491  &    -7.032  &    -7.072  \\
     2.000  &     2.420  &     0.991  &    -0.160  &    -1.837  &    -4.897  &    -6.073  &    -6.568  &    -6.535  \\
     3.000  &     2.632  &     1.171  &     0.045  &    -1.563  &    -4.603  &    -5.795  &    -6.317  &    -6.305  \\
     4.000  &     2.893  &     1.321  &     0.201  &    -1.360  &    -4.433  &    -5.640  &    -6.174  &    -6.169  \\
     5.000  &     2.948  &     1.411  &     0.307  &    -1.218  &    -4.285  &    -5.499  &    -6.034  &    -6.032  \\
     6.000  &     2.513  &     1.413  &     0.384  &    -1.076  &    -4.138  &    -5.358  &    -5.885  &    -5.880  \\
     7.000  &     2.745  &     1.540  &     0.490  &    -0.958  &    -4.036  &    -5.262  &    -5.785  &    -5.773  \\
     8.000  &     2.694  &     1.554  &     0.519  &    -0.912  &    -4.025  &    -5.248  &    -5.765  &    -5.744  \\
     9.000  &     2.844  &     1.635  &     0.593  &    -0.838  &    -3.995  &    -5.220  &    -5.734  &    -5.705  \\
    10.000  &     2.786  &     1.640  &     0.611  &    -0.815  &    -3.960  &    -5.181  &    -5.692  &    -5.663  \\
    11.000  &     2.783  &     1.685  &     0.664  &    -0.767  &    -3.917  &    -5.139  &    -5.650  &    -5.620  \\
    12.000  &     2.716  &     1.703  &     0.696  &    -0.732  &    -3.880  &    -5.102  &    -5.613  &    -5.582  \\
    13.000  &     2.818  &     1.744  &     0.729  &    -0.704  &    -3.856  &    -5.077  &    -5.587  &    -5.556  \\
    13.500  &     2.788  &     1.743  &     0.734  &    -0.695  &    -3.846  &    -5.066  &    -5.576  &    -5.545  \\
\hline
\end{tabular}
\end{scriptsize}
\end{minipage}
{(This table is available in its entirety in the online journal, and at CDS in machine-readable format.
Values for solar metallicity and helium content are shown here for guidance regarding its form and content.)}
\label{tabla3}
\end{table*}

\begin{table*}
 \centering
 \begin{minipage}{280mm}
  \caption{SBF amplitudes for CB09 models with fiducial mass-loss plus dusty envelopes}
  \begin{scriptsize}
  \begin{tabular}{@{}rrrrrrrrr@{}}
\hline
 Age (Gyr)  & $\barM_B$  & $\barM_V$  & $\barM_R$  & $\barM_I$  & $\barM_J$  & $\barM_H$  & $\barM_{K_s}$  & $\barM_K$ \\
\hline
\multicolumn {9}{|c|}{$Z = 0.017 \ \ Y = 0.26$}\\
\hline
     0.005  &    -7.637  &    -7.809  &    -8.061  &    -8.340  &    -8.999  &    -9.510  &    -9.644  &    -9.617  \\
     0.006  &    -7.446  &    -7.697  &    -8.168  &    -8.678  &    -9.705  &   -10.389  &   -10.559  &   -10.521  \\
     0.007  &    -7.171  &    -7.415  &    -7.887  &    -8.432  &    -9.546  &   -10.286  &   -10.487  &   -10.446  \\
     0.008  &    -6.885  &    -7.052  &    -7.457  &    -8.033  &    -9.295  &   -10.137  &   -10.390  &   -10.347  \\
     0.009  &    -6.427  &    -6.470  &    -6.701  &    -7.330  &    -8.979  &    -9.998  &   -10.330  &   -10.286  \\
     0.010  &    -6.096  &    -6.092  &    -6.352  &    -7.155  &    -8.963  &    -9.984  &   -10.313  &   -10.270  \\
     0.020  &    -4.936  &    -4.972  &    -5.263  &    -6.051  &    -7.692  &    -8.698  &    -8.999  &    -8.954  \\
     0.030  &    -4.390  &    -4.500  &    -4.818  &    -5.491  &    -6.951  &    -7.945  &    -8.232  &    -8.186  \\
     0.040  &    -3.884  &    -4.049  &    -4.370  &    -4.973  &    -6.338  &    -7.338  &    -7.621  &    -7.573  \\
     0.050  &    -3.382  &    -3.658  &    -4.031  &    -4.624  &    -5.937  &    -6.929  &    -7.205  &    -7.156  \\
     0.060  &    -2.990  &    -3.367  &    -3.795  &    -4.406  &    -5.737  &    -6.742  &    -7.027  &    -6.977  \\
     0.070  &    -2.251  &    -2.832  &    -3.385  &    -4.052  &    -5.581  &    -6.665  &    -7.028  &    -6.975  \\
     0.080  &    -1.895  &    -2.547  &    -3.172  &    -3.883  &    -5.510  &    -6.626  &    -7.017  &    -6.964  \\
     0.090  &    -1.629  &    -2.341  &    -3.019  &    -3.754  &    -5.397  &    -6.517  &    -6.914  &    -6.860  \\
     0.100  &    -1.326  &    -2.066  &    -2.799  &    -3.577  &    -5.291  &    -6.428  &    -6.837  &    -6.782  \\
     0.200  &    -0.291  &    -0.679  &    -1.538  &    -2.995  &    -7.066  &    -8.424  &    -8.685  &    -8.739  \\
     0.300  &     0.084  &    -0.265  &    -1.170  &    -2.763  &    -6.233  &    -7.724  &    -8.144  &    -8.223  \\
     0.400  &     0.510  &     0.093  &    -0.805  &    -2.477  &    -5.646  &    -7.198  &    -7.676  &    -7.766  \\
     0.500  &     0.816  &     0.310  &    -0.584  &    -2.277  &    -5.443  &    -6.961  &    -7.349  &    -7.421  \\
     0.600  &     0.999  &     0.410  &    -0.517  &    -2.208  &    -5.324  &    -6.852  &    -7.265  &    -7.341  \\
     0.700  &     1.382  &     0.623  &    -0.360  &    -2.024  &    -5.028  &    -6.603  &    -7.089  &    -7.179  \\
     0.800  &     1.582  &     0.727  &    -0.274  &    -1.898  &    -4.839  &    -6.457  &    -6.997  &    -7.096  \\
     0.900  &     1.765  &     0.802  &    -0.216  &    -1.800  &    -4.702  &    -6.346  &    -6.913  &    -7.013  \\
     1.000  &     1.949  &     0.876  &    -0.152  &    -1.708  &    -4.601  &    -6.245  &    -6.801  &    -6.890  \\
     1.500  &     2.091  &     0.443  &    -0.750  &    -2.336  &    -4.957  &    -6.392  &    -6.875  &    -6.950  \\
     2.000  &     2.426  &     0.986  &    -0.211  &    -1.620  &    -5.062  &    -6.408  &    -6.671  &    -6.702  \\
     3.000  &     2.635  &     1.169  &     0.033  &    -1.428  &    -4.818  &    -6.175  &    -6.445  &    -6.459  \\
     4.000  &     2.895  &     1.319  &     0.197  &    -1.326  &    -4.841  &    -6.219  &    -6.460  &    -6.454  \\
     5.000  &     2.950  &     1.410  &     0.306  &    -1.186  &    -4.658  &    -6.035  &    -6.300  &    -6.292  \\
     6.000  &     2.513  &     1.413  &     0.384  &    -1.040  &    -4.393  &    -5.754  &    -6.056  &    -6.047  \\
     7.000  &     2.745  &     1.540  &     0.491  &    -0.920  &    -4.204  &    -5.548  &    -5.882  &    -5.870  \\
     8.000  &     2.694  &     1.554  &     0.519  &    -0.875  &    -4.145  &    -5.473  &    -5.816  &    -5.802  \\
     9.000  &     2.844  &     1.635  &     0.593  &    -0.804  &    -4.081  &    -5.400  &    -5.755  &    -5.735  \\
    10.000  &     2.786  &     1.640  &     0.611  &    -0.784  &    -4.057  &    -5.369  &    -5.730  &    -5.708  \\
    11.000  &     2.783  &     1.685  &     0.664  &    -0.738  &    -4.022  &    -5.333  &    -5.701  &    -5.677  \\
    12.000  &     2.716  &     1.703  &     0.697  &    -0.706  &    -3.992  &    -5.300  &    -5.676  &    -5.650  \\
    13.000  &     2.818  &     1.744  &     0.729  &    -0.681  &    -3.970  &    -5.274  &    -5.656  &    -5.630  \\
    13.500  &     2.788  &     1.743  &     0.734  &    -0.672  &    -3.958  &    -5.258  &    -5.645  &    -5.618  \\
\hline
\end{tabular}
\end{scriptsize}
\end{minipage}
{(This table is available in its entirety in the online journal, and at CDS in machine-readable format.
Values for solar metallicity and helium content are shown here for guidance regarding its form and content.)}
\label{tabla4}
\end{table*}

\section{Broad-band colours.} \label{broadband}

\subsection{Individual AGB stars.}

\begin{figure*}
\includegraphics[width=0.70\hsize,clip=]{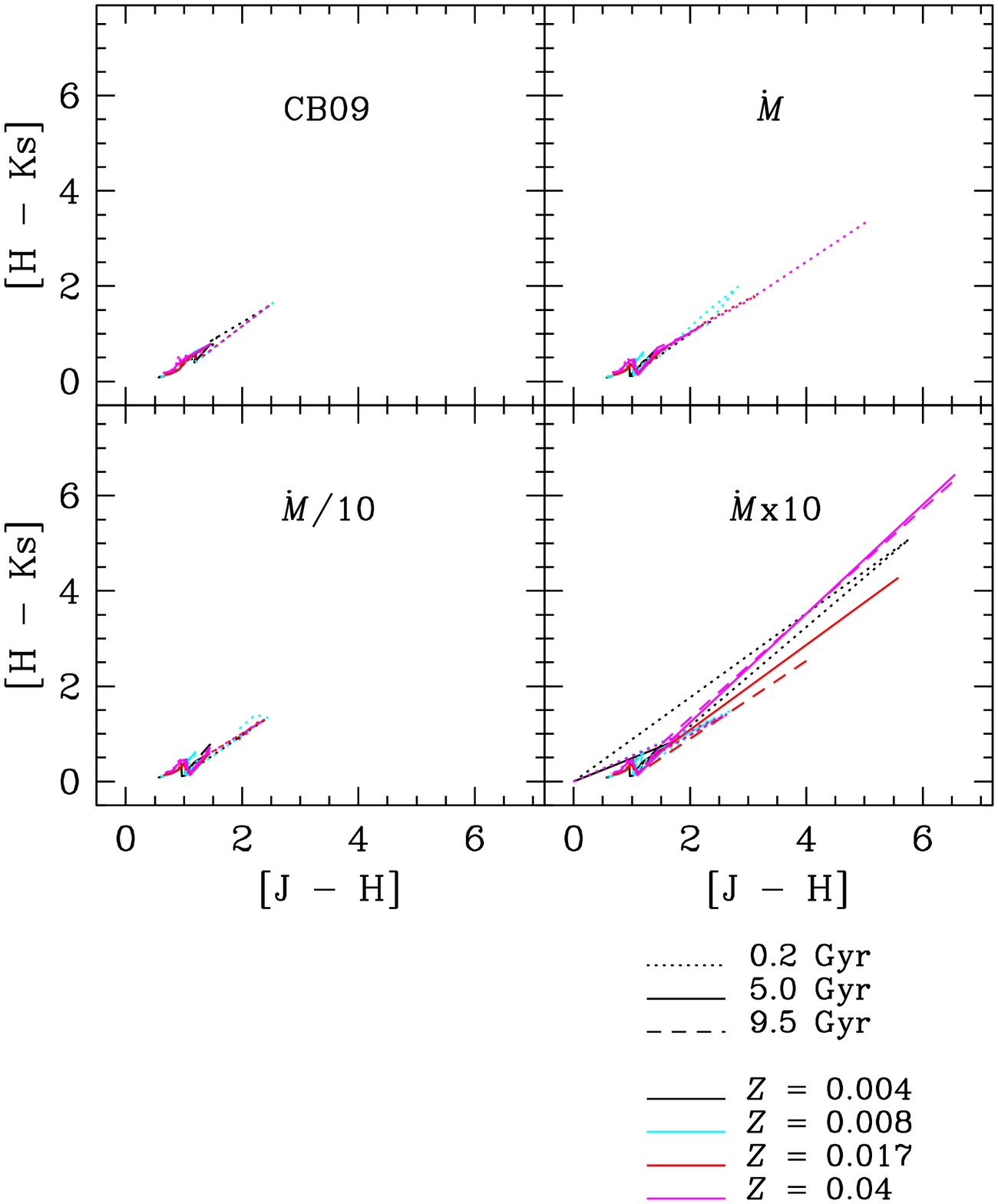}
\caption{
Theoretical two-colour diagrams, [$H -K_s$] vs.\ [$J- H$], of
individual stars along the 0.2 Gyr ({\it dotted line}), 5.0 Gyr ({\it
solid line}), and 9.5 Gyr ({\it dashed line}) isochrones, for
populations with different mass-loss rates and metallicities.  {\it
Top left:} standard CB09; {\it bottom left:} fiducial $\dot M / 10$;
{\it top right:} fiducial $\dot M$; {\it bottom right:} fiducial $\dot
M \times 10$. Different colours indicate diverse metallicities, i.e.,
{\it black:} $Z$ = 0.004; {\it cyan:} $Z$ = 0.008; {\it red:} $Z$ =
0.017; {\it magenta:} $Z$ = 0.04.  }
\label{piov_nir}
\end{figure*}

Fig.~\ref{piov_nir} displays the theoretical colours\footnote{We use $K_s$
or $K$, depending on the observations the models will be compared with.
The difference is very hardly noticeable.} of
individual stars along the 0.2 Gyr (dotted line), 0.5 Gyr (solid 
line), and 9.5 Gyr (dashed line) isochrones for populations with four metallicities
($Z =$ 0.004, black; 0.008, cyan; 0.017, red; and 0.04, magenta), and our four choices of spectra for stars in 
the TP-AGB. 

\begin{figure*}
\includegraphics[angle=-90.,width=0.85\hsize,clip=]{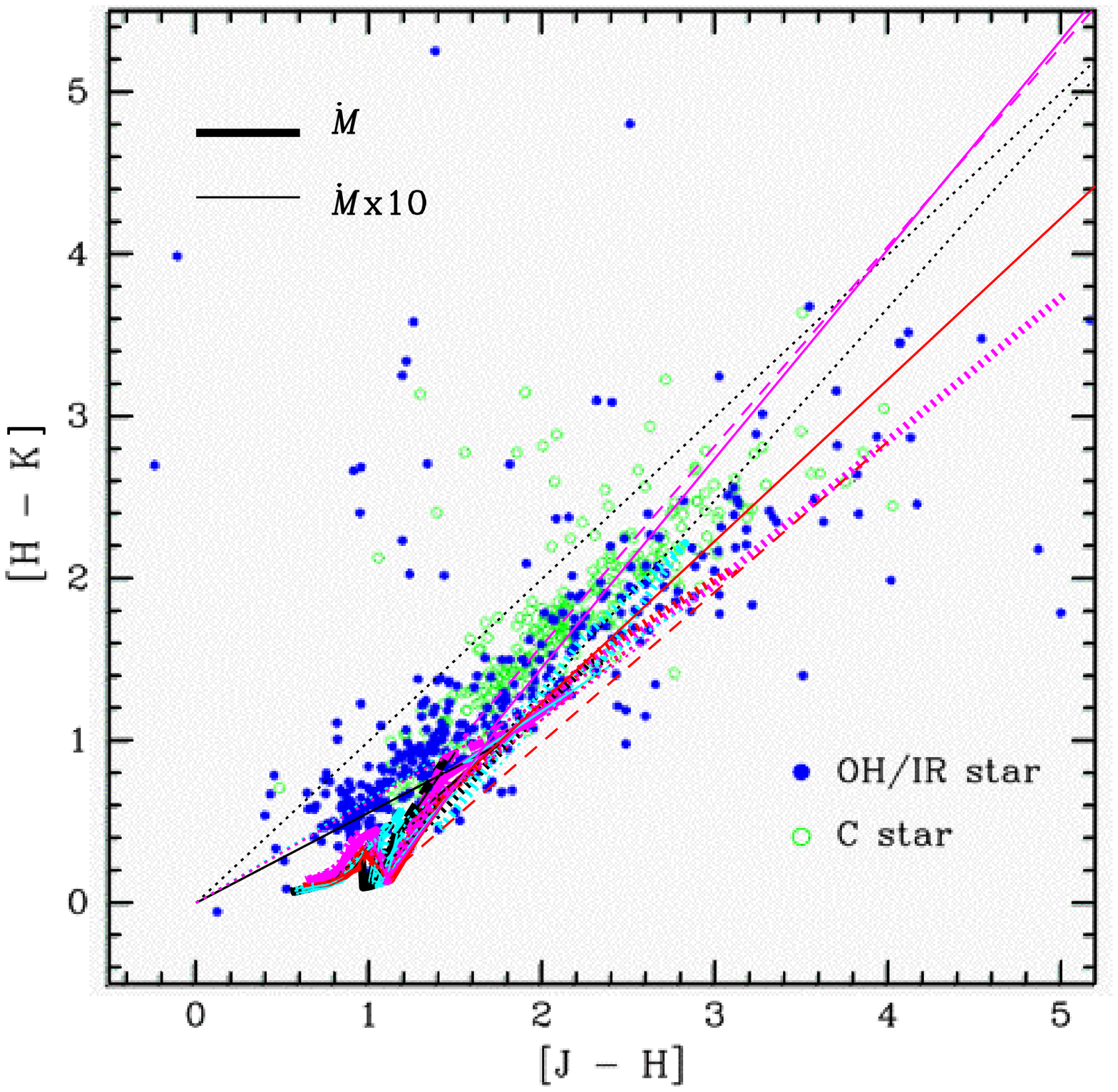}
\caption{
[$H -K$] vs.\ [$J- H$] integrated colours of individual OH/IR and Mira stars from
various sources in the literature. {\it Blue solid circles:} O-rich stars; 
{\it green open circles:} C-stars. The data were originally
compiled and properly corrected for extinction by 
\citet{piov03}. 
$J$, $H$, and $K$ magnitudes brighter than 7 have typical uncertainties of 0.05 mag. 
Theoretical colours of individual stars are 
superimposed on the sample of observed AGB stars. 
{\it Thick} lines: fiducial 
mass-loss rate $\dot M$; {\it thin} lines: high mass-loss rate 
$\dot M \times$ 10. Metallicity and age symbols as in Fig.\ \ref{piov_nir}. 
}
\label{overlay}
\end{figure*}

As a first test of the models, we compare the theoretical
near-infrared (near-IR) broad-band colours to the observed two-colour diagram,
[$H - K$] vs.\ [$J - H$], of individual AGB stars in the compilation
by \citet{piov03}; the sample is shown in Fig.\ \ref{overlay}.
Theoretical colours of
individual stars with fiducial (thick lines) and high 
(thin lines) mass-loss rates are also shown in Fig.~\ref{overlay}, superimposed on the observed
AGB sample.
The metallicity and age symbols are the same as in Fig.\ \ref{piov_nir}.

An examination of the bottom left panel of Fig.~\ref{piov_nir} (SEDs with $\dot M / 10$) 
illustrates that a population where TP-AGB stars are 
nearly devoid of dusty envelopes cannot explain the colour range of observed AGB stars.      
As for the standard CB09 isochrones (top left panel of Fig.~\ref{piov_nir}),   
not even the young populations with massive AGB stars reach beyond [$J - H$] $\sim$ 3 and 
[$H - K_s$] $\sim$ 2. 
By contrast, there are stars in the sample with  
both [$J - H$] and [$H - K_s$] $>$ 4. Models with fiducial $\dot M$ and dusty
envelopes (top right panel) cover these
values comfortably, but only at young ages and with solar or supersolar metallicities. 
Models with $\dot M \times 10$ (bottom right panel) and high metallicities fit the reddest stars  
regardless of age, whereas for lower $Z$ also young ages are required.

Figures~\ref{piov_nir} and~\ref{overlay} suggest that some of the stars in the sample 
with [$H - K_s$] $\geq$ 2 could be 
young ($\sim$ 0.2 Gyr old), comparatively massive
($M > 4 M_\odot$ at the beginning of the TP-AGB phase), stars
with subsolar metallicity,
going through a superwind phase with a mass-loss rate 
of $\sim$ a few $\times 10^{-5} - 10^4 M_\odot$ yr$^{-1}$, or
roughly one order of magnitude above the fiducial rate. 
Stars could continue losing mass at these rates for $\sim 10^7$ years.

\begin{figure*}
\includegraphics[width=0.85\hsize,clip=]{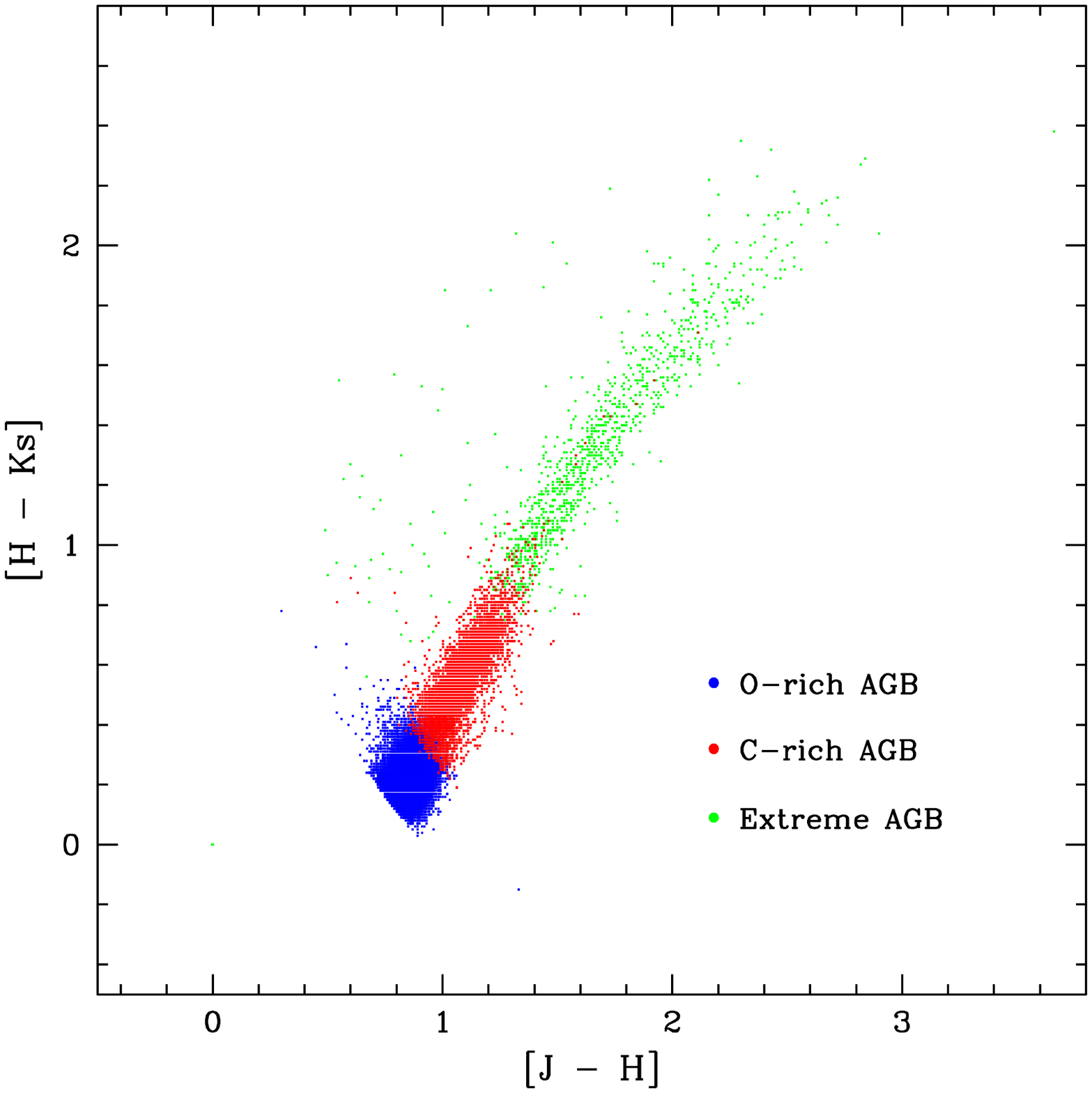}
\caption{
2MASS [$H - K_s$] vs.\ [$J - H$] integrated colours of individual AGB
candidates in the sample of \citet{srin09}. {\it Blue:} O-rich
stars. {\it Red:} C-rich stars.  {\it Green:} ``extreme" AGB stars;
these are the most luminous AGB stars, losing the most mass.  Typical
photometric errors are 0.05 mag for sources with $K_s \sim$ 11 mag.  }
\label{sundar_nir_2color}
\end{figure*}

\begin{figure*}
\includegraphics[width=0.85\hsize,clip=]{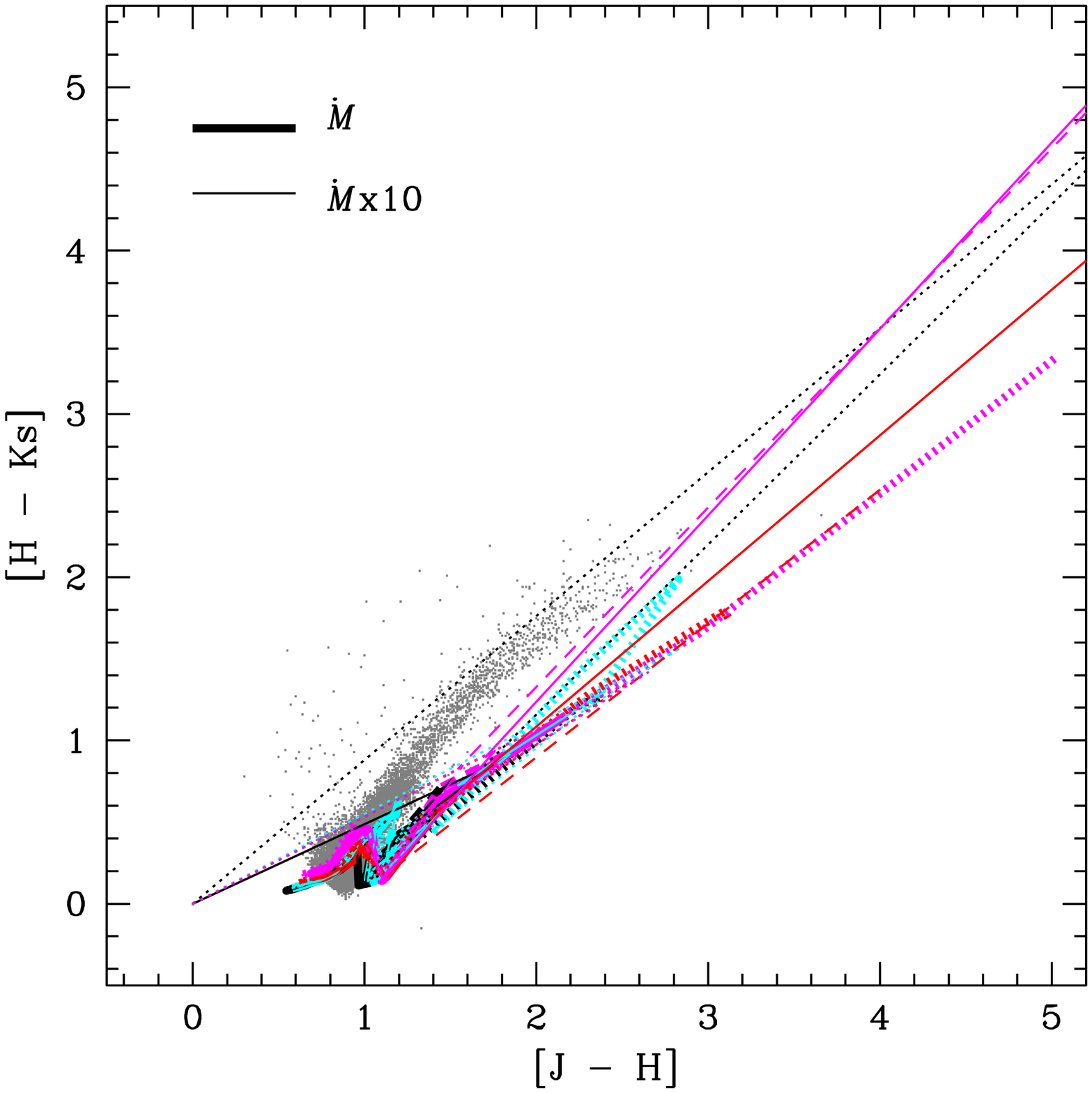}
\caption{Theoretical colours of individual stars are superimposed 
on the sample of observed AGB stars, now shown
as gray points. {\it Thick}
lines: fiducial mass-loss rate $\dot M$; {\it thin} lines: high
mass-loss rate $\dot M \times$ 10. Metallicity and age symbols as in
Fig.\ \ref{piov_nir}.
}
\label{sundar_nir_overlay}
\end{figure*}

For a more recent example, the 2MASS [$H - K_s$] vs.\ [$J - K_s$] integrated colours of individual AGB candidates in the
sample published by \citet{srin09} are displayed in 
Figure~\ref{sundar_nir_2color}. Different colours are used for O-rich, C-rich, and ``extreme" 
(based on their 2MASS and IRAC colours) AGB objects. The colour range is smaller than
that in Figure~\ref{overlay}. This can be better appreciated in 
Figure~\ref{sundar_nir_overlay}; now in the same scale as Figure~\ref{overlay}, the \citeauthor{srin09}
sample is shown as a cloud of gray points, with our models superimposed. Ages and metallicities
of the models are indicated as in Figures~\ref{piov_nir} and~\ref{overlay}. The agreement with the data is similar to 
that achieved for the older sample compiled by \citet{piov03}.    

\subsection{Star clusters.}

\begin{figure*}
\includegraphics[width=0.70\hsize,clip=]{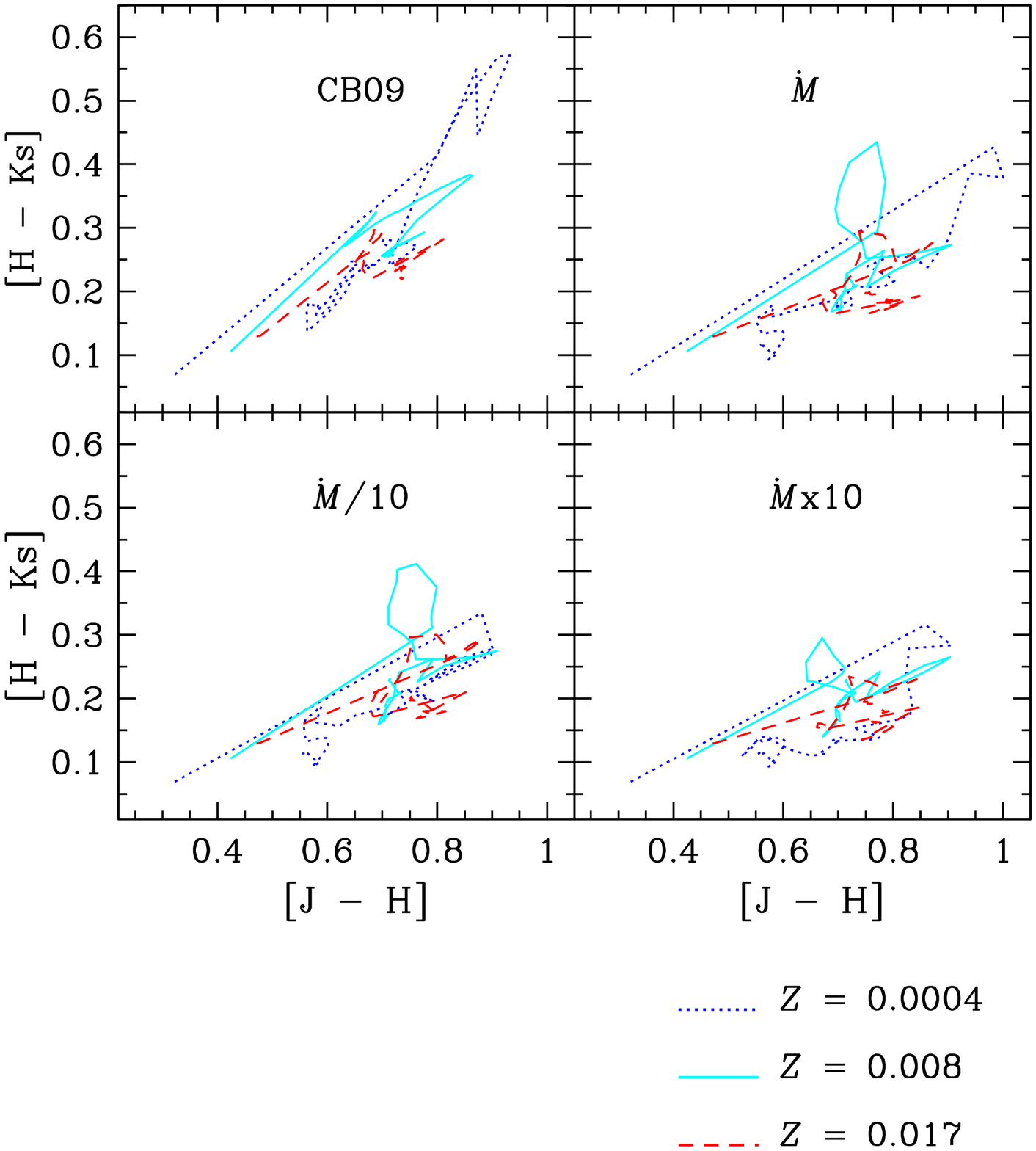}
\caption{Theoretical two-colour diagrams, [$H - K_s$] vs.\ [$J - H$],
for SSPs with different metallicities and mass-loss rates.
{\it Top left:} standard CB09; {\it bottom left:} fiducial $\dot M / 10$;
{\it top right:} fiducial $\dot M$; {\it bottom right:}
fiducial $\dot M \times 10$.
Models span an age range between 100 Myr and 14 Gyr.
{\it Blue:} $Z$ = 0.0004; {\it cyan:} 
$Z$ = 0.008; {\it red:} $Z$ = 0.017. 
}
\label{piov_jhk}
\end{figure*}

Fig.~\ref{piov_jhk} presents theoretical two-colour diagrams, [$H -
K_s$] vs.\ [$J - H$], for SSPs with different metallicities ($Z$ =
0.0004, blue; 0.008, cyan; 0.017, red) and, again, our 4 choices of
mass-loss and spectra for stars in the TP-AGB.
The model ages go from 100 Myr to 14 Gyr. 
The first thing that stands out is the range covered by the colour values.
As opposed to the colours of individual stars (see
Figures~\ref{piov_nir},~\ref{overlay}, and~\ref{sundar_nir_overlay}), 
the integrated colours of SSPs 
are in general confined to 
the very small range 0.3 $<$ [$J - H$] $<$ 1.0, 0.1 $\leq$ [$H - K_s$] $<$ 0.6, 
with [$J - H$] varying slightly more than [$H - K_s$], no matter what mass-loss rate 
is used for the models.

Still, it should be useful to compare the theoretical integrated
broad-band colours to those of stellar clusters. Our first test set
comprises the MC clusters measured by \citet{gonz04,gonz05a}.  These
authors assembled 8 artificial ``superclusters", by coadding data of
191 star clusters in bins with similar ages and metallicities,
according to classes I $-$ VII in the \citet[]{sear80} SWB
categorization scheme, plus an ultra-young (pre-SWB class)
supercluster.\footnote{Individual
clusters of each SWB class were centred, sky subtracted, 
multiplicatively scaled to a common photometric
zero-point and dereddened before coaddition. SMC
clusters were magnified to place them at the distance modulus of the
LMC. 
}
The purpose of such procedure is to reduce the
stochastic uncertainty produced by the inadequate sampling, in sparse
clusters, of stars evolving through short evolutionary phases, of
which the AGB is a prime example.  ``Superclusters", therefore, should
be more appropriate test objects than individual star clusters
\citep[see, for example,][]{sant97,bruz02,cerv02}.  In fact, if the
assumption is made that the numbers of stars in different evolutionary
stages have a Poissonian distribution, then the theoretical relative
errors of integrated colours scale as $M_{\rm tot}^{-1/2}$, where $M_{\rm tot}$
is the total mass of the stellar population \citep{cerv02}. In what
follows, we will use coloured regions to represent expected $\pm 1 \sigma$
stochastic errors.  The MC ``supercluster" ages, that go from $\sim$ 6
Myr to $>$ 10 Gyr, are not the originally adopted by
\citet{gonz04}. Instead, we use now the updated calibration by
\citet{gira95} of the $S$-parameter developed by
\citet{elso85,elso88}; this parameter relates the ages of LMC clusters
to their $UBV$ colours.  To each supercluster we assign the age that
corresponds to the ``central'' $S$-type of its constituents; the error
is set to span the $S$-types of all the members, plus and minus the
rms dispersion $\delta$ (log $t$) = 0.14 found by \citet{gira95} for the
log $t$ -- $S$ relation.

\begin{figure*}
\includegraphics[width=0.70\hsize,clip=]{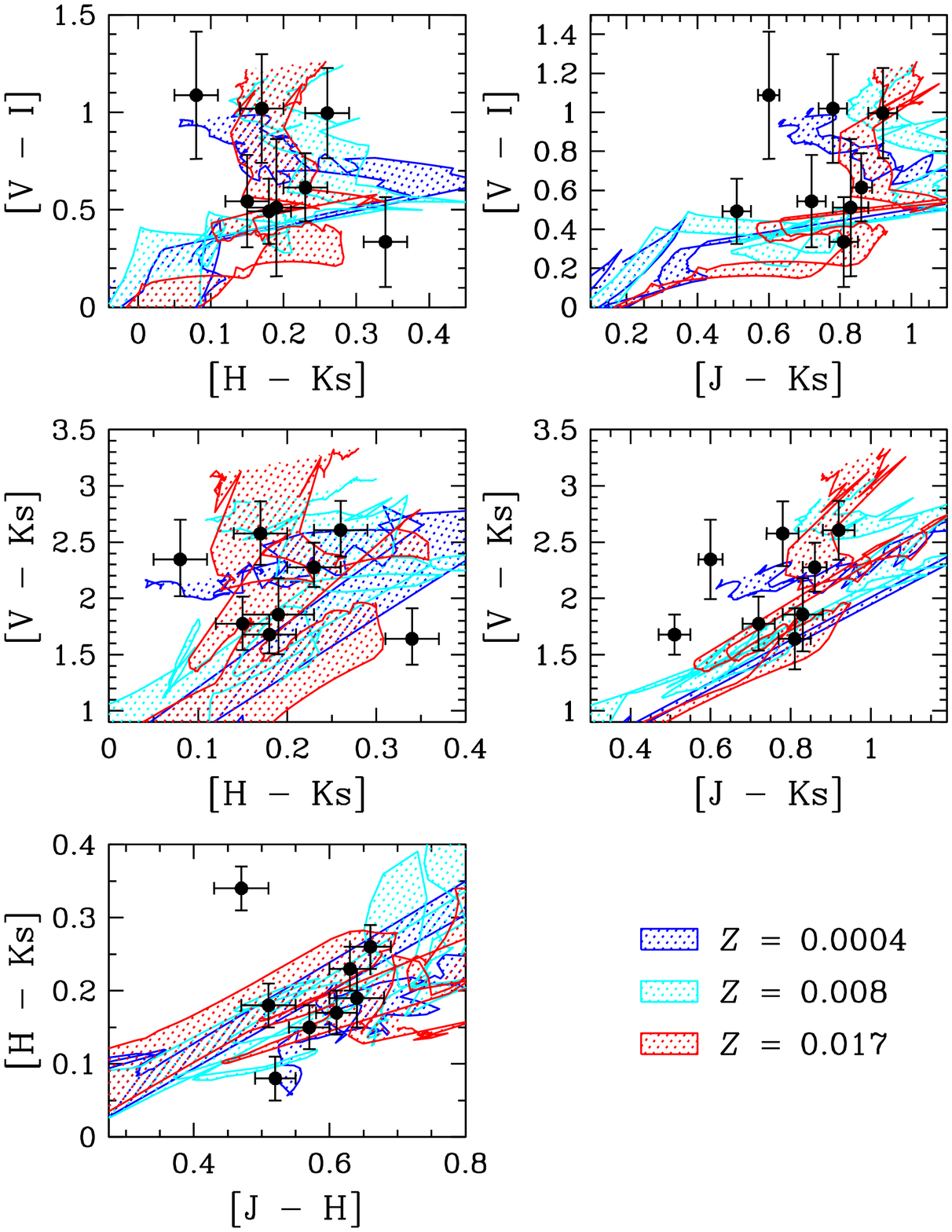}
\caption{Comparison between models and MC
``superclusters": two-colour diagrams.  {\it Top left:} [$V - I$] vs.\ [$H - K_s$]; {\it
top right:} [$V - I$] vs.\ [$J - K_s$]; {\it middle left:} [$V - K_s$]
vs.\ [$H - K_s$]; {\it middle right:} [$V - K_s$] vs.\ [$J - K_s$];
{\it bottom left:} [$H - K_s$] vs.\ [$J - H$]. Filled circles are
artificial ``superclusters" built by
\citet{gonz04,gonz05b,gonz05a}. Coloured regions represent SSPs with
fiducial $\dot M$ and expected $\pm 1 \sigma$ error bars for 5$\times
10^5~M_\odot$.  {\it Blue:} $Z =$ 0.0004; {\it cyan:} $Z =$ 0.008;
{\it red:} $Z =$ 0.017. Supercluster ages go from 6 Myr to 10 Gyr;
model ages span between 3 Myr and 14 Gyr.  }
\label{piov_sclust}
\end{figure*}

Figure \ref{piov_sclust} shows five two-colour diagrams, comparing model 
SSPs with data of the superclusters, respectively
[$V - I$] vs.\ [$H - Ks$], [$V - I$] vs.\ [$J - Ks$], 
[$V - Ks$] vs.\ [$H - Ks$], [$V - Ks$] vs.\ [$J - Ks$], 
and [$H - Ks$] vs.\ [$J - H$]. 
Near-IR data have been taken from the Two Micron All Sky Survey \citep[2MASS,][]{skru97}; 
$I$-band data were retrieved from the Deep Near-Infrared Southern Sky Survey
\citep[DENIS;][]{epch97}; and $V$ data come from different sources in
the literature, mostly from the compilation by 
\citet[][ see Gonz\'alez-L\'opezlira et al.\ 2005]{vand81}.
Near-IR colours for the superclusters were derived for the first time by 
\citet{gonz04}, [$V - I$] by \citet{gonz05a}, and [$V - K_s$] specifically for the
present work. 
We have rederived [$V - I$] and the near-IR colours, however, to make sure that the individual cluster
centres are right, that both an image and stellar photometry 
are available for all clusters included in each supercluster, and that the background 
subtraction is optimal.\footnote{We have discovered,
for example, that NGC~1854 and NGC~1855, both reportedly type SWB II, are actually the same cluster! 
Their putative centres are listed to be $6\farcs1$ apart; their respective $S$-parameters are, 
according to \citet{elso85}, 
24 and 22, a difference that provides an independent estimate of the uncertainty in the 
assignment of $S$.
}
Whereas the near-IR colours were measured in the supercluster mosaics, using
circular apertures with $r = 1\arcmin$, colours involving $V$ ([$V - I$], [$V - J$], [$V - H$], and [$V - K_s$]) 
were first obtained for single clusters, using the (diverse) diaphragms and $V$ magnitudes from
the compilation by \citet{vand81}, and then averaged to derive colours for the superclusters. 
In these cases, the quoted errors equal the dispersion of the individual colours, divided by
$(N-1)^{1/2}$, with $N$ the number of objects in each supercluster.
Measured colours for all superclusters are presented in Table 5, 
together with their ages, metallicities, and photometric masses. 
Except for supercluster SWB I, masses are always
lower than estimated by \citet{gonz04}; for types III--VII, this is due more to our younger (on average, half as old) adopted ages than to
changes between the BC03 (used by those authors) and the CB09 models. While the new assumed ages of types pre-SWB, I, and II  
are older (on average, twice as old), type I's estimated mass is the same and type II's, about 30\% lighter. 
The only noteworthy case is the pre-SWB supercluster: its new estimated mass is only 10\% of the one derived
by \citet{gonz04}, and hence its SBF amplitudes are shown now with
significantly larger uncertainties.  

In all the panels, the data (solid black circles with error bars) are compared to models with a fiducial mass-loss rate 
and 3 different metallicities ($Z$ = 0.0004, blue; 0.008, cyan; 0.017, red), that  
bracket those of the superclusters (0.0007 $ \leq Z \leq $ = 0.01; Frogel et al.\ 1990, assuming
that $Z_\odot = 0.017$; 
Cohen 1982). The expected $\pm 1 \sigma$ error bars for the models, shown as coloured bands, 
have been calculated as in Gonz\'alez et al.\ (2004, Appendix), assuming a stellar population
of $5 \times 10^5~M_\odot$. This value is conservative, and representative of the 
MC superclusters; they have between $\sim 10^5$ and $\sim 3\times 10^6~M_\odot$. The two superclusters
where the effects of AGB stars should be more important (aged 160 and 450 Myr, 
respectively) have both $\sim 5\times 10^5~M_\odot$ in stars.
Roughly, the models have no problem explaining the colours of the superclusters,
although we notice that the pre-SWB supercluster lies quite far away
from the models in the bottom left panel of the figure ([$H - K_s$] vs.\ [$J - H$]).
We hypothesise that this supercluster might suffer from 
additional extinction \citep[see][]{gonz05a} and systematic effects. 
Firstly, in the optical and even in the 
near-IR, there might be a selection against very young clusters that are the 
product of the most energetic star formation. In fact, although 
2MASS data for NGC 2070 (30 Dor) were available, they were not used 
to build the supercluster type pre-SWB because they showed abundant  
nebular emission and dust extinction. Also, regarding very young clusters,
the assumption that the addition of many small objects is statistically equal to a 
large one will fail, if none of the small clusters is massive enough
to produce the most massive stars. This problem is not relevant after
a few $10^7$ yr, when these stars die and cease to contribute to the
cluster's light.

\begin{figure*}
\includegraphics[width=0.70\hsize,clip=]{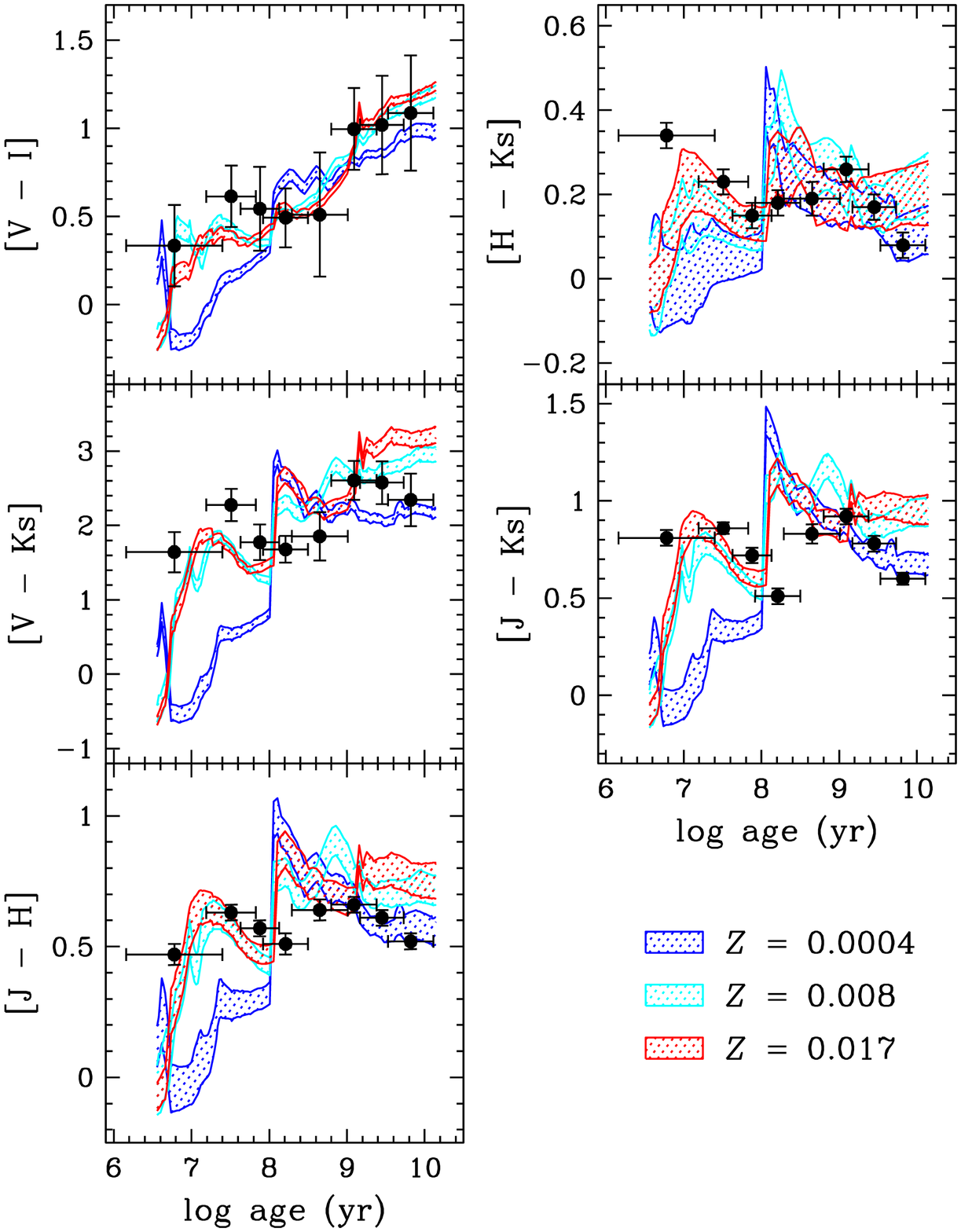}
\caption{Comparison between models and MC
``superclusters": colours vs.~log (age). {\it Top left:} [$V - I$]; {\it middle left:} [$V -
K_s$]; {\it bottom left:} [$J - H$]; {\it top right:} [$H - K_s$];
{\it middle right:} [$J - K_s$].  Coloured regions represent SSPs with
fiducial $\dot M$ and expected $\pm 1 \sigma$ error bars for 5$\times
10^5~M_\odot$, coded as in Figure~\ref{piov_sclust}.}
\label{age_cols}
\end{figure*}

Perhaps more useful to assess the models is the comparison with the
data in the age--colour planes.  Figure~\ref{age_cols} shows the data
compared to colour versus age for models with a fiducial mass-loss
rate. The trends of colour with age shown by the data are very closely
followed by the models.

\begin{figure*}
\includegraphics[width=0.85\hsize,clip=]{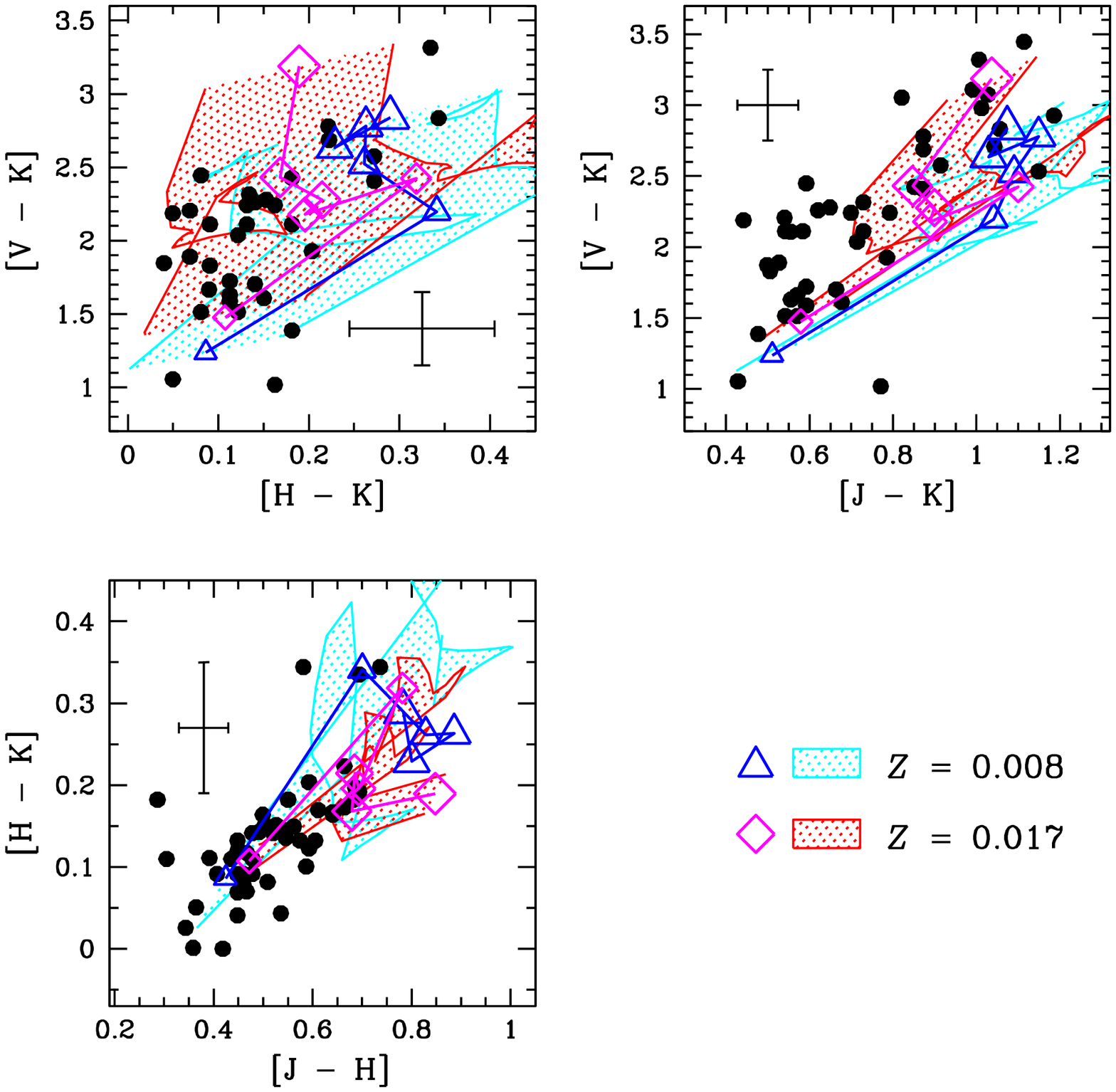}
\caption{Comparison between models and Magellanic star clusters. {\it
Top left:} [$V - K$] vs.\ [$H - K$]; {\it top right:} [$V - K$] vs.\
[$J - K$]; {\it bottom left:} [$H - K$] vs.\ [$J - H$]. Filled circles
are clusters compiled from the literature and reddening-corrected by
\citet{piov03}. Coloured regions represent SSPs with fiducial $\dot M$
and expected  $\pm 1 \sigma$ error bars for $10^5 M_\odot$. {\it Cyan:} $Z
=$ 0.008 ({\it cyan}); {\it red:} $Z =$ 0.017.  Most of the clusters
are younger than 1.5 Gyr; the models range in age between 100 Myr and
1.5 Gyr. Blue triangles and magenta diamonds indicate 0.1, 0.3, 0.5,
0.1, 1.1, and 1.5 Gyr, respectively for $Z = 0.008$ and $Z = 0.017$;
increasing symbol size represents increasing age.  }
\label{piov_clust}
\end{figure*}

It is also instructive to compare our models to the cluster sample already compiled and presented by 
\citet{piov03}.\footnote{
The data were extracted by us from Piovan et al.'s paper with 
Dexter \citep{deml01}.} 
Once again, two-colour diagrams are shown in Fig.\ \ref{piov_clust}:  
[$V - K$] vs.\ [$H - K$]; [$V - K$] vs.\ [$J - K$]; and
[$H - K$] vs.\ [$J - H$]. The data 
are clusters mostly younger than 1.5 Gyr, and are displayed as filled circles,
with average error bars shown for each panel. The expected $\pm 1 \sigma$ error bars for the models 
are depicted once more as coloured regions, except that now we are assuming a population of 
$10^5~M_\odot$. 
The models have a fiducial mass-loss rate; and metallicities 
$Z =$ 0.008 ({\it cyan}) and 
$Z =$ 0.017 ({\it red}), that encompass those of the clusters. Model ages go from 
100 Myr to 1.5 Gyr. 

The match is comparable to that achieved by Piovan et al.\ (2003, their Figure 16) for most
colours, and considerably
better for [$V - K$] vs.\ [$H - K$], even though our model ages in the figure stop at 
1.5 Gyr (theirs go up to 15 Gyr).

We have seen so far that our models are able to fit both near-IR colours of most single AGB stars, 
and integrated optical and 
near-IR colours of star clusters with 
different ages and metallicities. Integrated colours of star clusters, however, do not
seem to be able to discriminate between different choices of global mass-loss rates.  
Next, we will investigate whether surface brightness fluctuations are potentially
sensitive to different mass-loss rates in stellar populations.

\section{Surface brightness fluctuations, metallicity, and mass-loss.} \label{sec_theosbf}

The technique of surface brightness fluctuation (SBF) measurements 
was introduced by \citet{tonr88} as a way to 
derive distances to early-type galaxies. The fluctuation flux 
(denoted \barf) is the ratio between the variance and the mean
of the stellar luminosity function \citep{tonr88,tonr90}, scaled by 
$(4 \pi d^2)^{-1}$, where $d$ is the distance. This is 
expressed as follows: 

\begin{equation}
\barf = \frac {1}{4 \pi d^2} \frac{\Sigma  n_i l_i^2}{\Sigma n_i l_i}; 
\label{theequation}
\end{equation} 

\noindent $n_i$ and $l_i$ are, 
respectively, the number of stars of type $i$, and their luminosity.  

In the case of galaxies, the fluctuation magnitude\footnote{
\barm\ = $-$2.5 log \barf\ + zero point.} \barm\ 
is measured through the spatial fluctuations 
in their surface brightness, 
and the distance is found by comparing \barm\ with 
empirically calibrated relations that give the absolute fluctuation 
magnitude, \barM, in a photometric band as a function of a certain 
broadband colour, in a given range \citep[e.g.,][]{wort93a,wort93b,pahr94,tonr97,ajha97,liu00,
liu02,mei01,jens03,cant03,cant05,mei05a,mei05b,gonz05a,mari06,mei07,cant07a,cant07b,blak09}.
In nearby stellar clusters, 
it is possible to obtain \barm\ by performing the sums 
in equation \ref{theequation} over resolved 
stars \citep{ajha94,gonz04,gonz05a,mouh05,raim05}. 

It is not hard to see that SBFs convey information 
about stellar populations, akin to
integrated photometry and spectra. However, because of their 
dependence on the square of the stellar luminosity, they 
are especially sensitive to, and can 
provide additional information about the brightest stars 
at a particular wavelength and at a given evolutionary
phase of a stellar population.
Accordingly, it has been suggested recently 
(Cantiello et al.\ 2003; 
Raimondo et al.\ 2005, Gonz\'alez-L\'opezlira \& Buzzoni 2008) that 
SBFs can be used to study AGB stars
in intermediate-age populations and, specifically, to 
investigate their mass-loss rates. 
These works, though, 
do not explore a possible intrinsic connection between metallicity 
and mass-loss, 
nor consider the impact of extinction by 
dust in the stellar envelope on the detectability of mass-losing
stars.

Surprisingly in a way, the relation between metallicity and 
mass-loss turns out to be controversial, even for the dust-driven
winds in the TP-AGB. On the one hand, detailed theoretical models 
\citep[e.g.][]{will00} predict that mass-loss should increase 
with metallicity, and \citet{groe95} have found, from fits to
$8 - 13 \mu$m spectra, mass-loss rate  
ratios of 4:3:1 for three O-rich AGB stars with similar periods 
in, respectively, our Galaxy, the LMC, 
and the Small Magellanic Cloud (SMC).\footnote{The present-day [Fe/H] ratios for the Sun, and B-type
stars in the LMC and the SMC are $\sim$ 3:2:1 \citep{moki07}; according to 
\citet{lyub05}, the Sun and B-type MS stars in
the solar neighbourhood have the same metallicity. 
The ratios for the Sun and F-type stars are $\sim$ 4:2:1 \citep{russ89}.} More recently, \citet{kali05,kali08} 
have found evidence of a metallicity dependence of the initial-final
mass relationship (between the mass of a white dwarf remnant and its main-sequence 
predecessor) from spectroscopic observations of 
white dwarfs in open clusters.
On the other hand, \citet{gail86} propose that the mass-loss rate
is proportional to the ratio $\tau/v_{\rm exp}$, where $\tau$ is the
optical depth of a dust-driven wind and $v_{\rm exp}$ is its 
velocity. Van Loon (2000) derives a metallicity-independent
mass-loss rate for a sample of dust-obscured C and O-rich AGB stars, also in
the Milky Way, the LMC, and the SMC. Correspondingly, \citet{vanl06} 
argues that both $\tau$ and $v_{\rm exp}$ 
depend on the square root of the dust-to-gas ratio, $\Psi$, 
that presumably is itself linearly proportional to metallicity,
such that the dependence of $\dot M$ on $Z$ cancels out. 
Today, however, we know that $\Psi$ is not constant with metallicity, and that both the dust-to-gas
ratio and the dust species in the stellar envelopes vary during evolution for a single
star and between different stars with the same initial metallicity 
\citep[see, e.g.,][]{lebz06,ferrarotti06}.


With the aim of addressing the question of the relation between 
$Z$ and $\dot M$, we compute the time evolution of SBF magnitudes 
of single-burst stellar populations in the $B$, $V$, $R$, $I$, $J$, $H$, and $K_S$ bands, 
in the nine metallicities and helium contents mentioned before.
The model fluctuation luminosity \barL\ at each wavelength is calculated with the
following equation (very similar to eq.~\ref{theequation}):

\begin{equation}
\barL = \frac{\Sigma w_i l_i^2}{\Sigma w_i l_i},
\label{theosbf}
\end{equation}

\noindent
where the weight $w_i$ is the number of stars of type $i$ {\it per unit mass} in the
population (set as explained in Section~\ref{dotmnstarpars}), and $l_i$ 
is the luminosity of stellar type $i$.\footnote{The corrected 
stellar weights were, of course, also applied to the calculation of integrated colours used 
in Section~\ref{broadband}.}

Following \citet{cerv02}, \citet{gonz04} demonstrated that the 
theoretical relative errors of fluctuation magnitudes and colours also
scale as $M_{\rm tot}^{-1/2}$, if a Poissonian distribution is assumed for 
the stellar numbers in different evolutionary phases.
In the rest of this paper, we show calculations for stellar populations with $5 \times 10^5 M_\odot$; 
this number is representative of the MC ``superclusters". 

\section{SBF Results.} \label{sbfres}

\subsection{Model SBFs.} \label{mresults}

\begin{figure*}
\includegraphics[width=0.85\hsize,clip=]{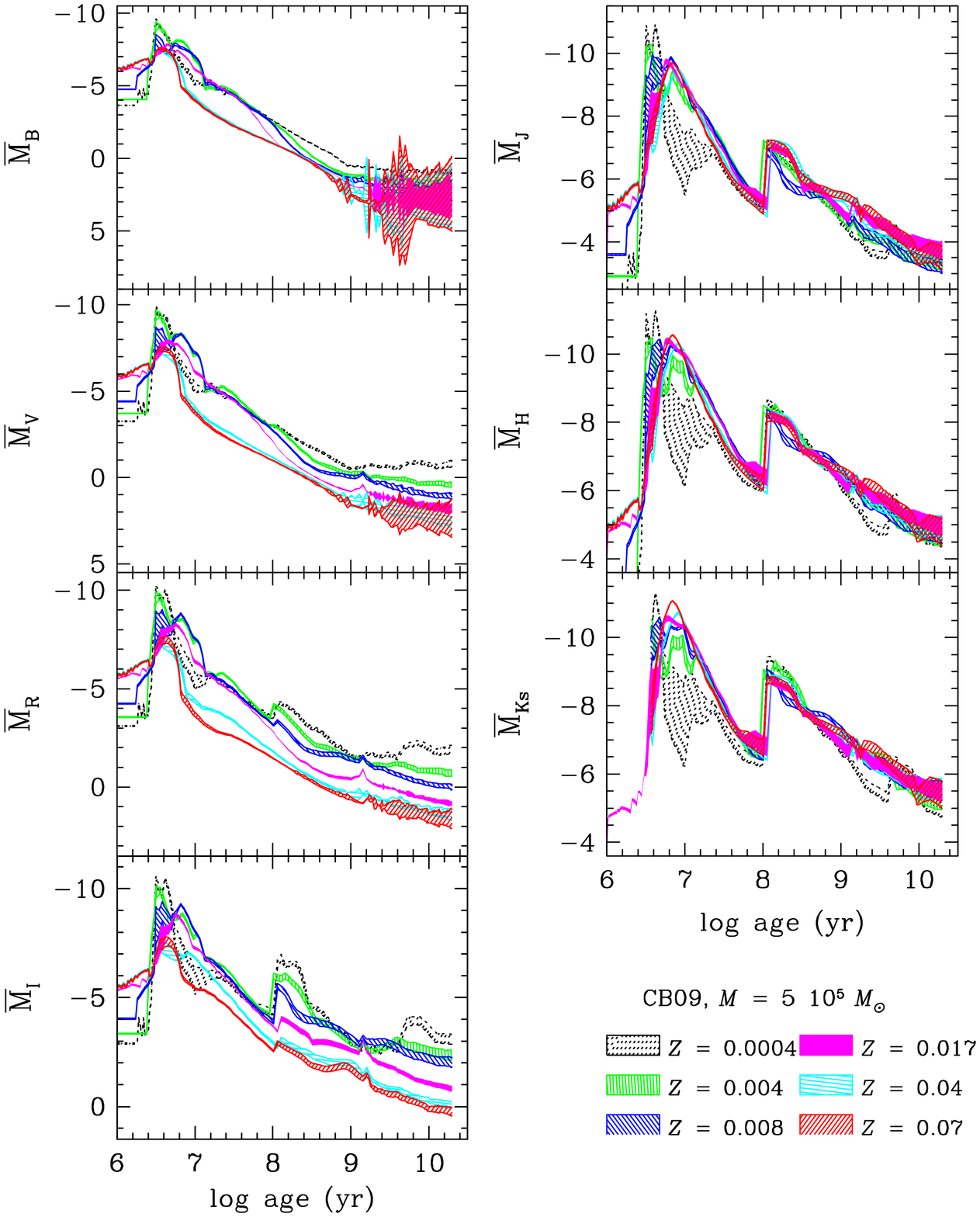}
\caption{Absolute fluctuation magnitudes vs.\ log (age) for standard
CB09 models with different metallicities.  Coloured regions delimit
expected  $\pm 1 \sigma$ stochastic errors for stellar populations with $5
\times 10^5~M_\odot$. {\it Black-dotted:} $Z$ = 0.0004; {\it
green-vertical-hatched:} $Z$ = 0.004; {\it blue-left-hatched:} $Z$ =
0.008; {\it magenta-solid:} $Z$ = 0.017; {\it
cyan-horizontal-hatched:} $Z$ = 0.04; {\it red-right-hatched:} $Z$ =
0.07.  }
\label{figcb}
\end{figure*}

Figure \ref{figcb} shows absolute fluctuation magnitudes vs.\ log (age) for standard CB09
models with different metallicities, from $Z$ = 0.0004 ($\sim$ 1/40 solar) to 
$Z =$ 0.07 ($\sim$ 4 times solar). 
Coloured regions delimit
expected $\pm$ 1 $\sigma$ stochastic errors for a stellar population with
$5 \times 10^5~M_\odot$. In the optical bands ($B$ to $I$), 
SBFs grow systematically fainter as metallicity increases.  
The reason is that the brightest stars (those that dominate the SBF signal) will be 
cooler (redder) at higher metallicities. Another consequence of the same phenomenon
is that errors grow notably larger with metallicity in
$B$ and $V$ for ages greater than a few Gyr; a  higher metallicity translates into 
fewer hot and blue stars, implying larger stochastic errors. 
A very similar trend with metallicity is seen for all the mass-loss rates explored in this paper.

The main difference with metallicity in the near-IR happens at very young ages ($\sim 10^7$ yr), 
between the most metal poor models ($Z = 0.0004$), and the rest. Metal poor populations will produce
SBFs about 2 mag fainter in the near-IR at these ages, as a consequence of having fewer red supergiants; for the
same reason, the stochastic error is largest ($\sim \pm 1$ mag) for these metallicities and ages. 
Although it is strictly true that SBFs increase with metallicity in the $J$ $H$, and $K_s$ bands
for populations $\sim$  100 -- 300 Myr old,
the sensitivity to $Z$ is smaller at $1 - 2 \mu$m. This is because, 
once the RGB is born, it is the most important contributor to  
the integrated light of stellar populations at these wavelengths
\citep[see, e.g.,][]{gonz04}; 
the variance (i.e., the SBFs) produced by the brightest red stars (both in the RGB and AGB) 
against this already bright background 
will then be less prominent.  
The onset of the TP-AGB is clearly discernible as a peak at about 10$^8$ years. 

\begin{figure*}
\includegraphics[width=0.70\hsize,clip=]{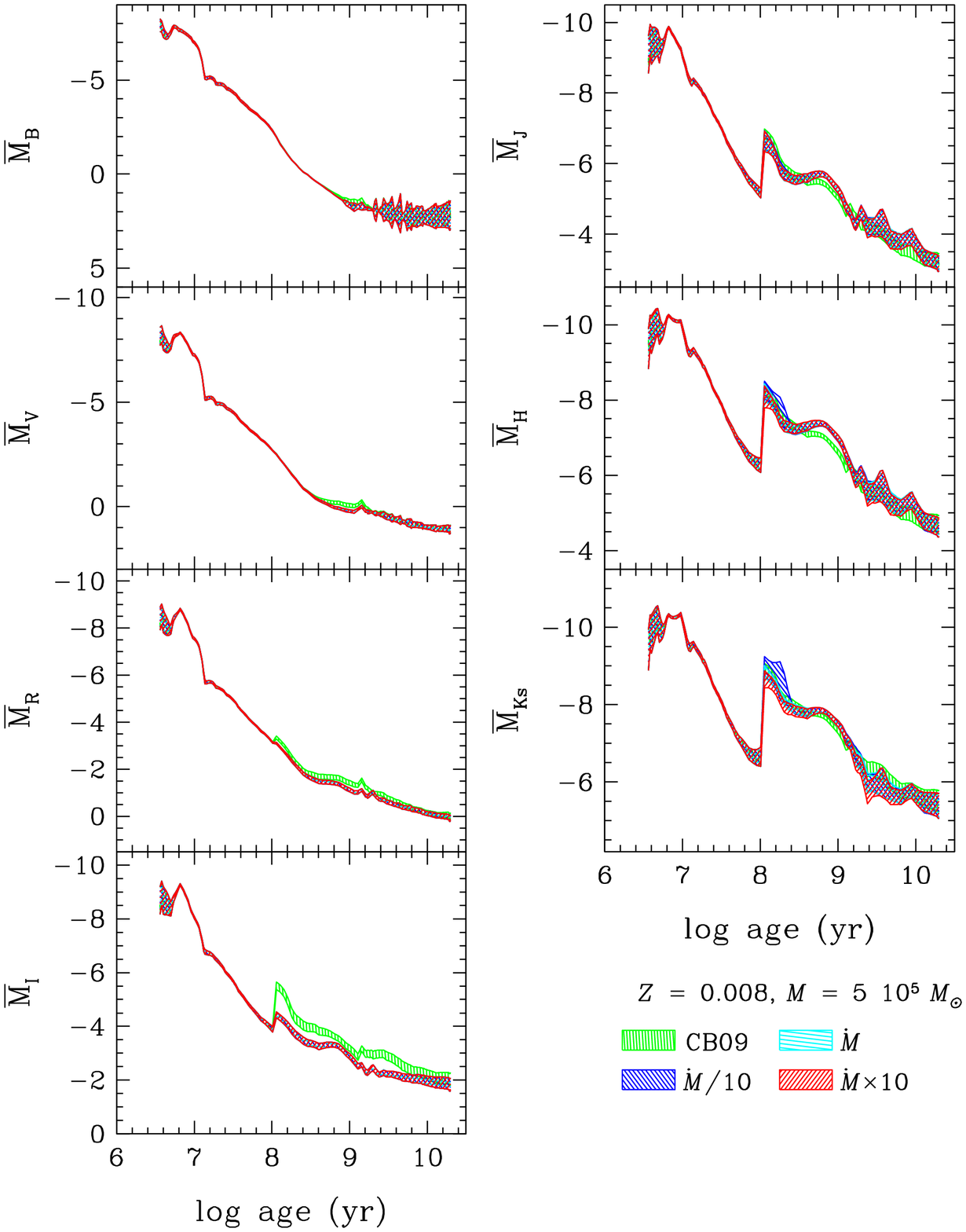}
\caption{Absolute fluctuation magnitudes vs.\ log (age) for $Z$ =
  0.008; models with different mass-loss rates.  Coloured regions
  delimit expected $\pm 1 \sigma$ stochastic errors for stellar
  populations with $5 \times 10^5~M_\odot$. 
  {\it Green-vertical-hatched:} standard CB09; {\it
    blue-left-hatched:} fiducial $\dot M$/10; {\it
    cyan-horizontal-hatched:} fiducial $\dot M$; {\it
    red-right-hatched:} fiducial $\dot M \times$10.
}
\label{figZ52}
\end{figure*}

Figure \ref{figZ52} displays absolute fluctuation magnitudes vs.\ log
(age) for models with $Z = 0.008$ and different mass-loss rates.  This
metallicity ($Z_\odot$/2.1) is closest to that of the youngest half of
the MC star clusters whose data are presented below.  Coloured regions
delimit expected $\pm 1 \sigma$ stochastic errors for a stellar population
with $5 \times 10^5 M_\odot$.  Contrary to the results of
\citet{raim05}, we find no strong trend of SBF brightness with mass-loss
rate. The biggest difference is between the CB09 models without dusty
envelopes and the models with dusty cocoons, regardless of $\dot M$.
This difference is most noticeable in the $I$-band, where models with
dust are about 0.5 mag fainter than CB09 standard models at all ages
after 10$^8$ yr.  The dusty models basically all fall on top of each
other.  Other metallicities show exactly the same behavior, with the
exceptions of (1) the already pointed out lower brightness and larger
dispersion of SBFs values for $Z = 0.0004$ and $Z = 0.001$ at 10$^7$
yr, and (2) a large scatter for the higher metallicities ($Z = 0.017$
to $Z = 0.07$) at ages greater than $\sim 2$ Gyr, in the $B$, $V$, and
$R$ passbands. In the latter case, the dispersion is caused by the
lower emission from red giants in these bands; it increases with
metallicity and decreases with wavelength, and goes from $\sim \pm
0.5$ mag, for $Z = 0.017$ at $V$, to $\sim \pm 4$ mag for $Z = 0.07$
at $B$.

The reason for this degeneracy is that the selection effects
highlighted by \citet{will00} are intensified by extinction, and
further exacerbated when the mass-loss rate is changed. The
probability of detecting the effects of stars shedding their envelopes
in an exponential fashion, already low owing to the short duration of
the phase, will decrease if the stars are heavily dust-enshrouded. If
the mass-loss-rate is modified upward, the intrinsic luminosity of
stars in the superwind stages will increase, but their lifetimes in
the phase will hence be even shorter and they will be more
obscured. The stars losing the most mass sometimes will not survive
past the first or second superwind stages.  Contrariwise, if mass-loss
rate is changed downward, stars will last longer and be less obscured,
but their luminosity will go correspondingly down.

\subsection{SBF data.} \label{sbfdat}

SBF magnitudes of MC star clusters have been determined previously by
\citet{gonz04,gonz05b} in the near-IR, and by \citet{raim05} in the
optical.  As discussed earlier, the former built 8 artificial
``superclusters",
whereas the latter chose to analyse Wide Field Planetary Camera 2 
(WFPC2) Hubble Space Telescope ({\sl HST}) $V$ and $I$ data of a dozen 
populous MC globular clusters.

\citeauthor{gonz04} derived SBFs 
for the ``superclusters" within a radius of 1$\arcmin$, as prescribed by equation
\ref{theequation}. The numerator was found by 
summing the square of the flux of resolved, bright stars,
obtained from the 2MASS Point Source Catalog;
field contamination was minimized by 
excluding from the analysis stars in the range
12.3 $< (K_s)_o <$ 14.3 with colours ($J - K_s$)$_o >$ 1.2 or
($J - K_s$)$_o <$ 0.4 \citep{ferr95}. 
Since the sum in the denominator converges slowly, it was computed from the total light
detected in the images, after removal of the emission in an annulus with
2$\farcm0 < r \leq 2\farcm5$; this was assumed to include the contributions 
from both sky and field stars.
Absolute fluctuation magnitudes 
were assigned taking an LMC distance modulus ${(m-M)_0 = 18.5}$ \citep{ferr00}.

As it turns out, the colour function used by \citet{gonz04} to select
cluster stars is adequate for 
older clusters, but is too red for the 2 or 3 youngest superclusters.
For this reason, we have recalculated near-IR fluctuation values for
all the superclusters.
This time, we have statistically removed the field population as per
the procedure described by \citet{migh96}.  
We compare, for each supercluster, the [$J - K_s$] 
versus $K_s$ diagram of stars 
within $r = 1\arcmin$ of the
supercluster centre (i.e., the ``cluster region"), 
with that of stars in an annulus with 2\farcm0 $< r \leq$
2\farcm5 (i.e., the ``field"); the former colour-magnitude diagram (CMD) presumably includes both cluster and field 
stars, whereas we assume that the latter
contains only field stars. For each star in the cluster region with 
mag $Ks \pm \sigma_{Ks}$ and colour [$J - K_s$] $\pm \sigma_{[J - Ks]}$,
we count the number of stars in the same CMD with [$J - K_s$] colours within
$\pm$MAX(2$\sigma_{[J - Ks]}$,0.100) mag and $K_s$ mag within 
$\pm$MAX(2$\sigma_{Ks}$,0.200) mag. We call this number $N_{\rm scl}$. 
We also count the number of stars in the field CMD within the same
$\Delta K_s$ by $\Delta$ [$J - K_s$] bin determined from the cluster star.
We call this number $N_{\rm fld}$. The probability $p$ that the star
in the cluster region CMD actually belongs to the supercluster 
can be expressed as:

\begin{equation}
p \approx 1 - {\rm MIN}\left( \frac{\alpha(N_{\rm
fld}+1)}{N_{\rm scl}+1},1.0\right),
\label{probdec}
\end{equation}

\noindent
where $\alpha$, in this case 0.44, is the ratio of the area of the cluster region ($\pi$ arcmin$^2$)
to the area of the field region (2.25 $\pi$ arcmin$^2$).
Once $p$ is calculated for a given star, it is compared
to a randomly drawn number $0 \leq p^\prime \leq 1$. If $p \geq p^\prime$, 
the star is
accepted as a supercluster member; otherwise, it is rejected and
considered as a field object. 

The numerator of equation \ref{theequation} was calculated with the
decontaminated star lists. 
We have obtained the denominator 
in the same way as before,
i.e., from the total light within $r = 1^\prime$, after subtracting
the light in the annulus between 2$\farcm0$ and 2$\farcm5$. 
The new SBF values tend to be fainter than 
the Gonz\'alez et al.\ (2004) ones; both sets of values agree within 1 $\sigma$, though, with the
exception of the SWB I supercluster, whose SBF determinations coincide 
within 3 $\sigma$.\footnote{
An alternative approach for the decontamination, using the field CMD as the reference (i.e., ``field" 
stars are removed from the cluster region, rather than ``cluster" stars kept), 
has been used and described by \citet{gall03}. We have tried this method,
and obtain the same results. We have also tried using the
annulus with 2\farcm5 $< r \leq$3\farcm0  as the ``field". Once again, the results do not vary
significantly, except for the pre-SWB supercluster, although 
even in this case both new derived sets of SBF values lie within the (considerably large) errors.  
}

We have also derived $\barM_I$ for the eight superclusters
using DENIS images, and photometry of the point sources from the DENIS database\footnote{Third release of DENIS data, 
The DENIS Consortium, 2005; http://cdsarc.u-strasbg.fr/viz-bin/Cat?B/denis.} 
and the DENIS Point Source Catalogue towards the Magellanic Clouds \citep{cion00}. 
The procedure is the same as for near-IR fluctuation
magnitudes, except that 
[$I - J$] versus $I$ diagrams have been employed for the  
field decontamination. SBF measurements for the superclusters are provided in Table 5. 

\begin{landscape}
\begin{table*}
   \hspace*{-5.8cm}
   \vspace*{5.0cm}
  \begin{minipage}{480mm}
  \begin{scriptsize}
   \caption{Characteristic parameters of Magellanic superclusters}
   \begin{tabular}{@{}lcccccccccccccc@{}}
   \hline
  Supercluster & Log age (yr){\normalsize $^a$} & $Z$  & Mass (10$^6 M_\odot$){\normalsize $^b$} &
[$V - I$] & [$V - J$] & [$V - H$] & [$V - K_s$] & [$J - Ks$] & [$H - 
K_s$] & [$J - K_s$] & $\barM_I$& $\barM_J$& $\barM_H$ & $\barM_{K_s}$ \\
  \hline
pre\dotfill &6.78$\pm$0.62 & 0.010$\pm$0.005{\normalsize $^c$} & 0.07 $\pm$ 0.02  &
0.34$\pm$0.23 &  0.89$\pm$0.26 &  1.32$\pm$0.27 &  1.64$\pm$0.27 &
0.47$\pm$0.18 &  0.34$\pm$0.18 &   0.81$\pm$0.18 &  -6.10$\pm$0.43 &
-5.86$\pm$0.64 &    -6.76$\pm$0.72  &   -6.78$\pm$0.73 \\
I\dotfill   &7.51$\pm$0.32 & 0.010$\pm$0.005{\normalsize $^{c}$} & 0.7 $\pm$ 0.1  &
0.62$\pm$0.18 &  1.40$\pm$0.21 &  2.04$\pm$0.21 &  2.28$\pm$0.22 &
0.63$\pm$0.14 &  0.23$\pm$0.14 &   0.86$\pm$0.14 &  -5.79$\pm$0.26 &
-6.77$\pm$0.20 &    -7.60$\pm$0.19  &   -7.77$\pm$0.19 \\
II\dotfill  &7.88$\pm$0.25 & 0.010$\pm$0.005{\normalsize $^{c}$} & 0.7 $\pm$ 0.1  &
0.54$\pm$0.24 &  1.10$\pm$0.24 &  1.62$\pm$0.24 &  1.78$\pm$0.24 &
0.57$\pm$0.13 &  0.15$\pm$0.12 &   0.72$\pm$0.13 &  -4.76$\pm$0.35 &
-6.59$\pm$0.54 &    -7.34$\pm$0.46  &   -7.58$\pm$0.41 \\
III\dotfill &8.21$\pm$0.29 & 0.010$\pm$0.005{\normalsize $^d$} & 0.4 $\pm$ 0.1   &
0.49$\pm$0.17 &  1.06$\pm$0.18 &  1.54$\pm$0.18 &  1.68$\pm$0.18 &
0.51$\pm$0.11 &  0.18$\pm$0.11 &   0.51$\pm$0.11 &  -3.03$\pm$0.20 &
-5.97$\pm$0.28 &    -7.11$\pm$0.27  &    -7.46$\pm$0.29 \\
IV\dotfill  &8.65$\pm$0.36 & 3e-3$\pm$2e-3{\normalsize $^{d}$} & 0.3 $\pm$ 0.0   &
0.51$\pm$0.35 &  1.13$\pm$0.33 &  1.65$\pm$0.33 &  1.86$\pm$0.32 &
0.64$\pm$0.16 &  0.19$\pm$0.15 &   0.83$\pm$0.15 &  -2.37$\pm$0.19 &
-5.43$\pm$0.26 &    -6.67$\pm$0.26  &   -7.12$\pm$0.25 \\
V\dotfill   &9.09$\pm$0.29 & 4e-3$\pm$2e-3{\normalsize $^{d}$} & 0.5 $\pm$ 0.1
& 1.00$\pm$0.23 &  1.77$\pm$0.26 &  2.38$\pm$0.26 &  2.61$\pm$0.26 &
0.66$\pm$0.16 &  0.26$\pm$0.16 &   0.92$\pm$0.16 &  -2.30$\pm$0.11 &
-4.32$\pm$0.16 &    -5.70$\pm$0.19  &   -6.46$\pm$0.23 \\
VI\dotfill  &9.45$\pm$0.28 & 2e-3$\pm$1e-3{\normalsize $^{d}$} & 1.8 $\pm$ 0.1
& 1.02$\pm$0.28 &  1.78$\pm$0.29 &  2.39$\pm$0.29 &  2.58$\pm$0.29 &
0.61$\pm$0.17 &  0.17$\pm$0.16 &   0.78$\pm$0.17 &  -1.74$\pm$0.18 &
-3.69$\pm$0.20 &    -5.04$\pm$0.23  &    -5.80$\pm$0.29 \\
VII\dotfill &9.82$\pm$0.29 & 7e-4$\pm$4e-4{\normalsize $^{d}$} & 2.8 $\pm$ 0.3
& 1.09$\pm$0.33 &  1.74$\pm$0.36 &  2.26$\pm$0.36 &  2.35$\pm$0.35 &
0.52$\pm$0.22 &  0.08$\pm$0.21 &   0.60$\pm$0.22 &  -2.65$\pm$0.35 &
-2.71$\pm$0.26 &    -3.91$\pm$0.36  &   -4.48$\pm$0.51 \\

\hline
\end{tabular}
\\

\end{scriptsize}
{$^a$ From the calibration of the $S$-parameter by \citet{gira95}.} \\
{$^b$ Masses from near-IR mass-to-light ratios, CB09 models; errors are
equal to the dispersion of the results at $J$, $H$, and $K_s$.} \\
{$^c$ \citet{cohe82}.} \\
{$^d$ \citet{frog90}, assuming $Z_\odot = 0.017$.}

\end{minipage}
\label{latabla}
\end{table*}
\end{landscape}

\begin{figure*}
\includegraphics[width=1.\hsize]{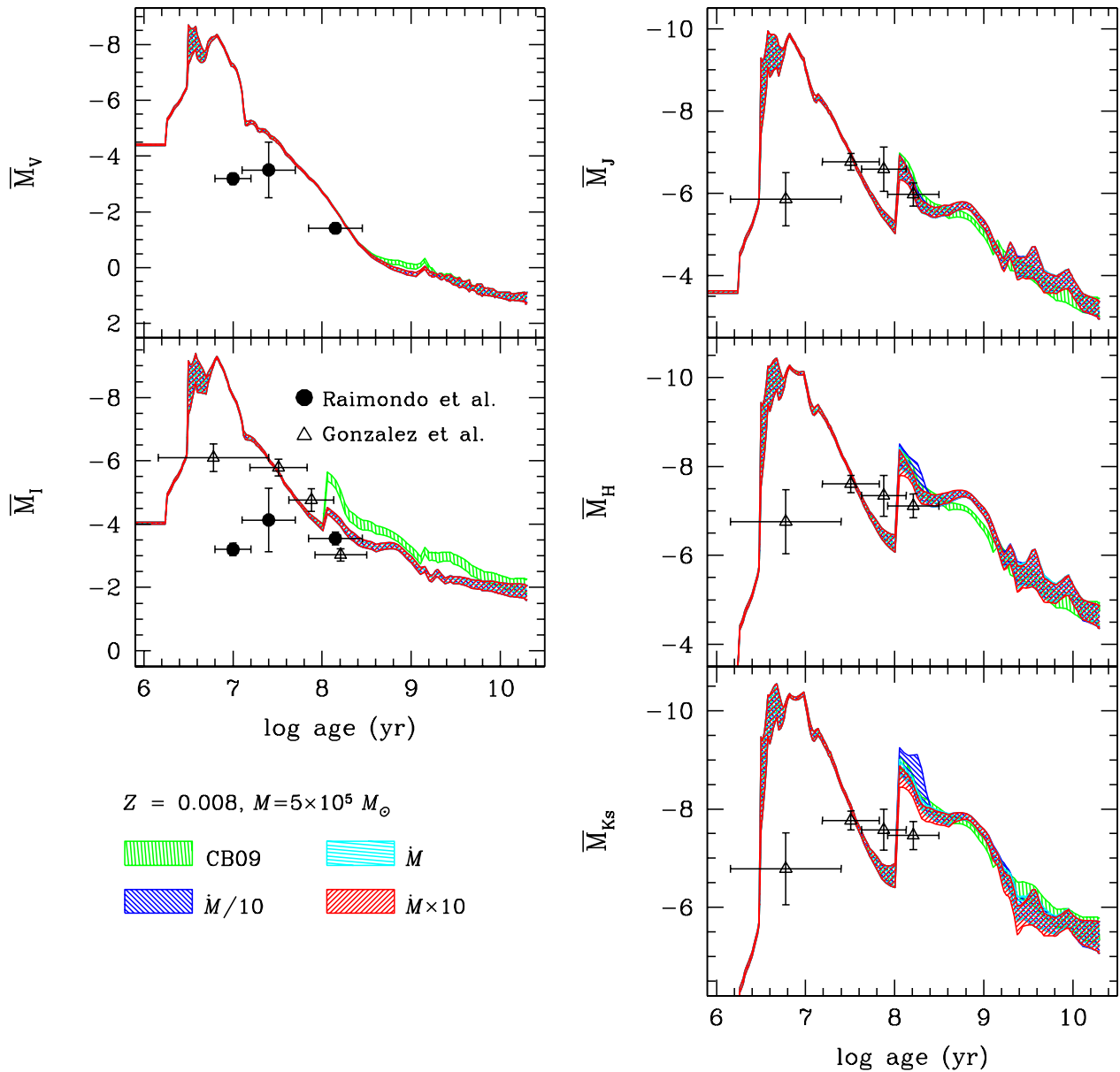}
\caption{SBFs of Magellanic star clusters I. Observations of young and
intermediate-age MC star clusters are compared to models with $Z$ =
0.008. $\pm 1 \sigma$ stochastic uncertainties are shown for stellar
populations with 5 $\times 10^5~M_\odot$, coded as in Figure
\ref{figZ52}. {\it Solid circles} are $\barM_V$ and $\barM_I$
measurements for globular clusters from \citet{raim05}; {\it empty
triangles} are $\barM_I$ and near-IR SBF measurements for artificial
MC ``superclusters" \citep[see][]{gonz04,gonz05b}.  }
\label{figclust52}
\end{figure*}

\begin{figure*}
\includegraphics[width=1.\hsize]{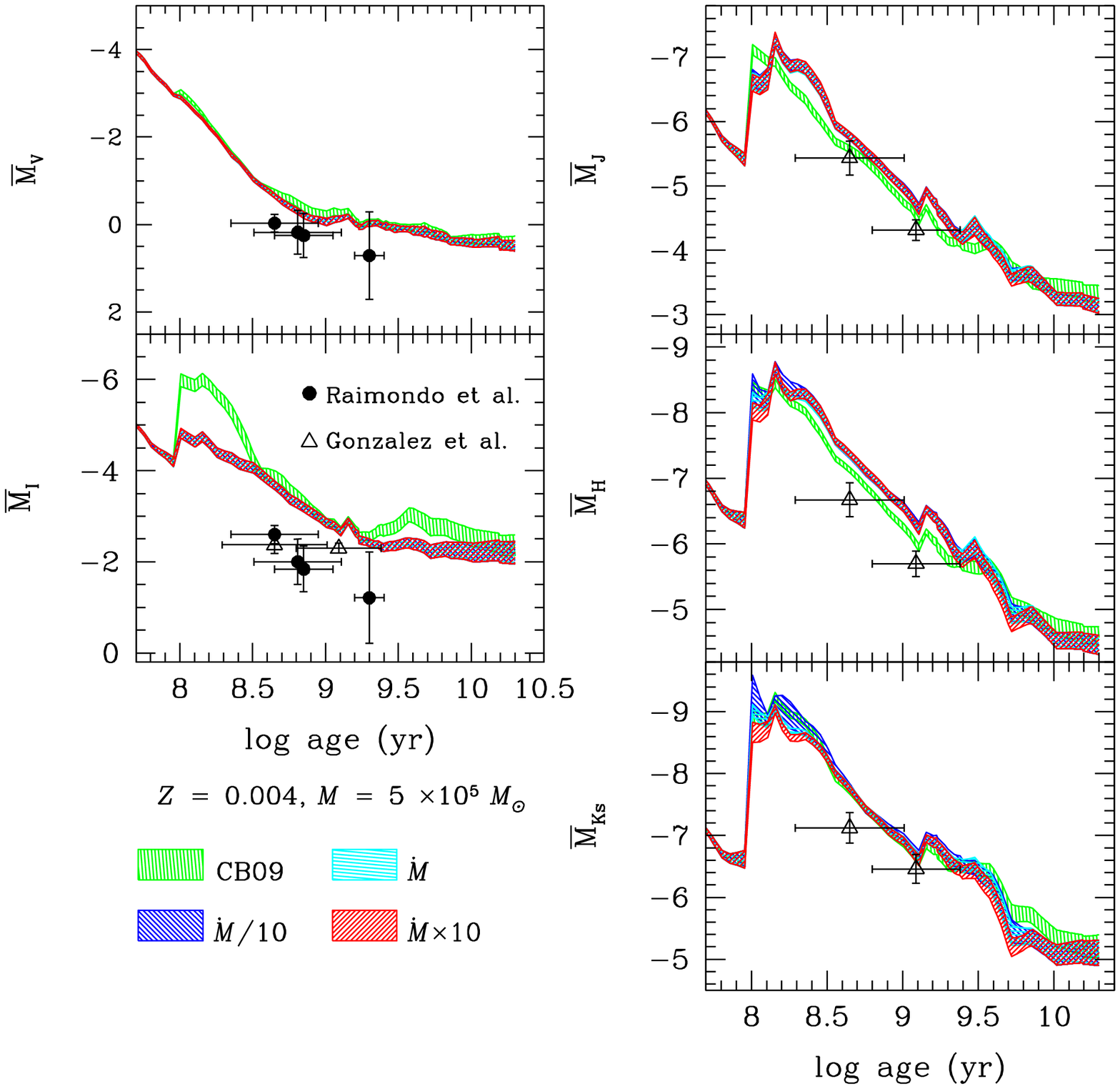}
\caption{SBFs of Magellanic star clusters II. Observations of
intermediate-age MC clusters are compared to models with $Z$ =
0.004. $\pm 1 \sigma$ stochastic uncertainties are shown for stellar
populations with $5 \times 10^5~M_\odot$, coded as in Figure
\ref{figZ52}.  }
\label{figclust42}
\end{figure*}

\begin{figure*}
\includegraphics[width=1.\hsize]{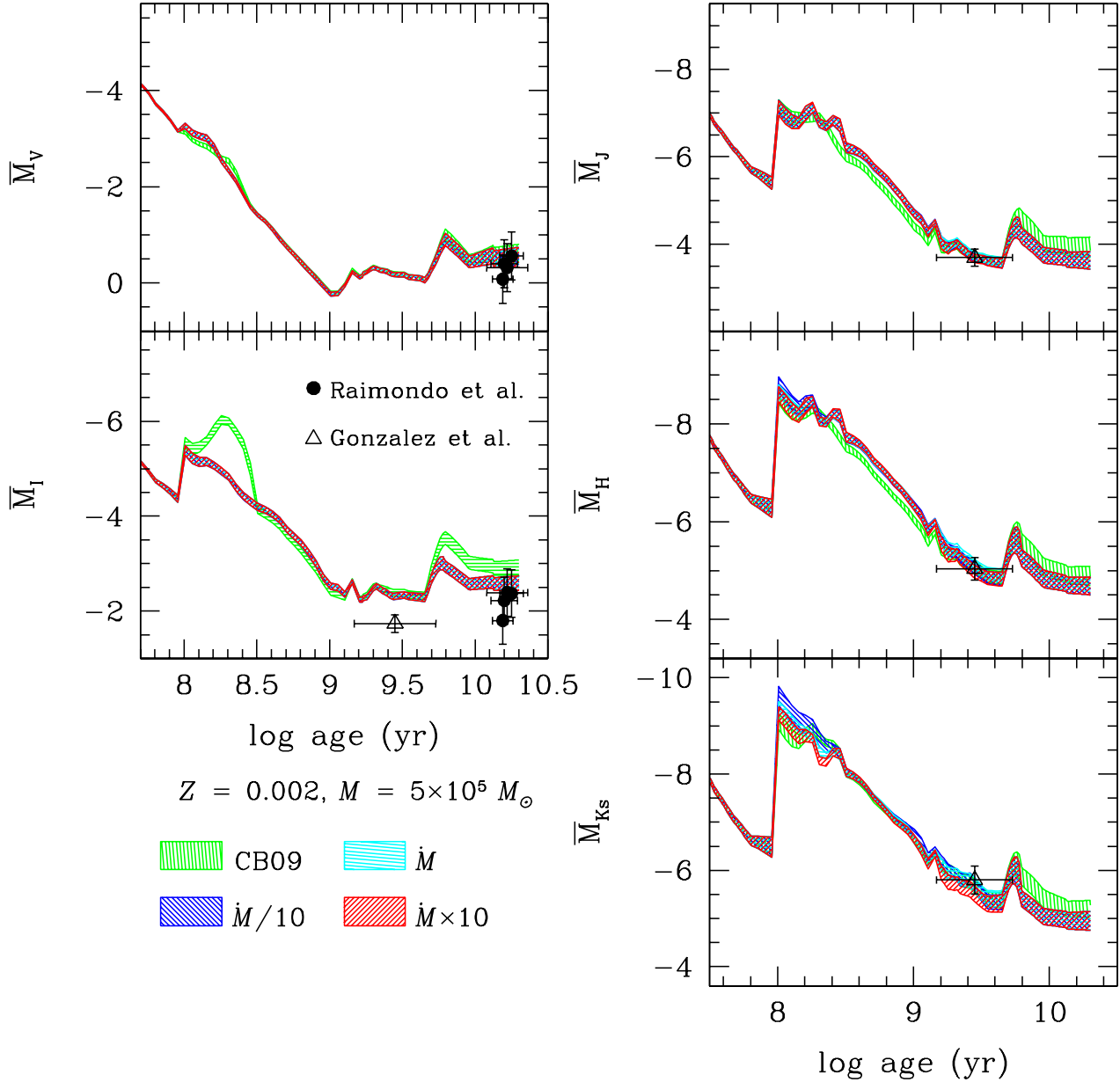}
\caption{SBFs of Magellanic star clusters III.  Observations of old MC
clusters are compared to models with $Z$ = 0.002. $\pm 1 \sigma$
stochastic uncertainties are shown for stellar populations with $5
\times 10^5~M_\odot$, coded as in Figure \ref{figZ52}.  According to
their reported SWB type (VI), Raimondo's four oldest MC clusters
belong here.  }
\label{figclust32}
\end{figure*}

\begin{figure*}
\includegraphics[width=1.\hsize]{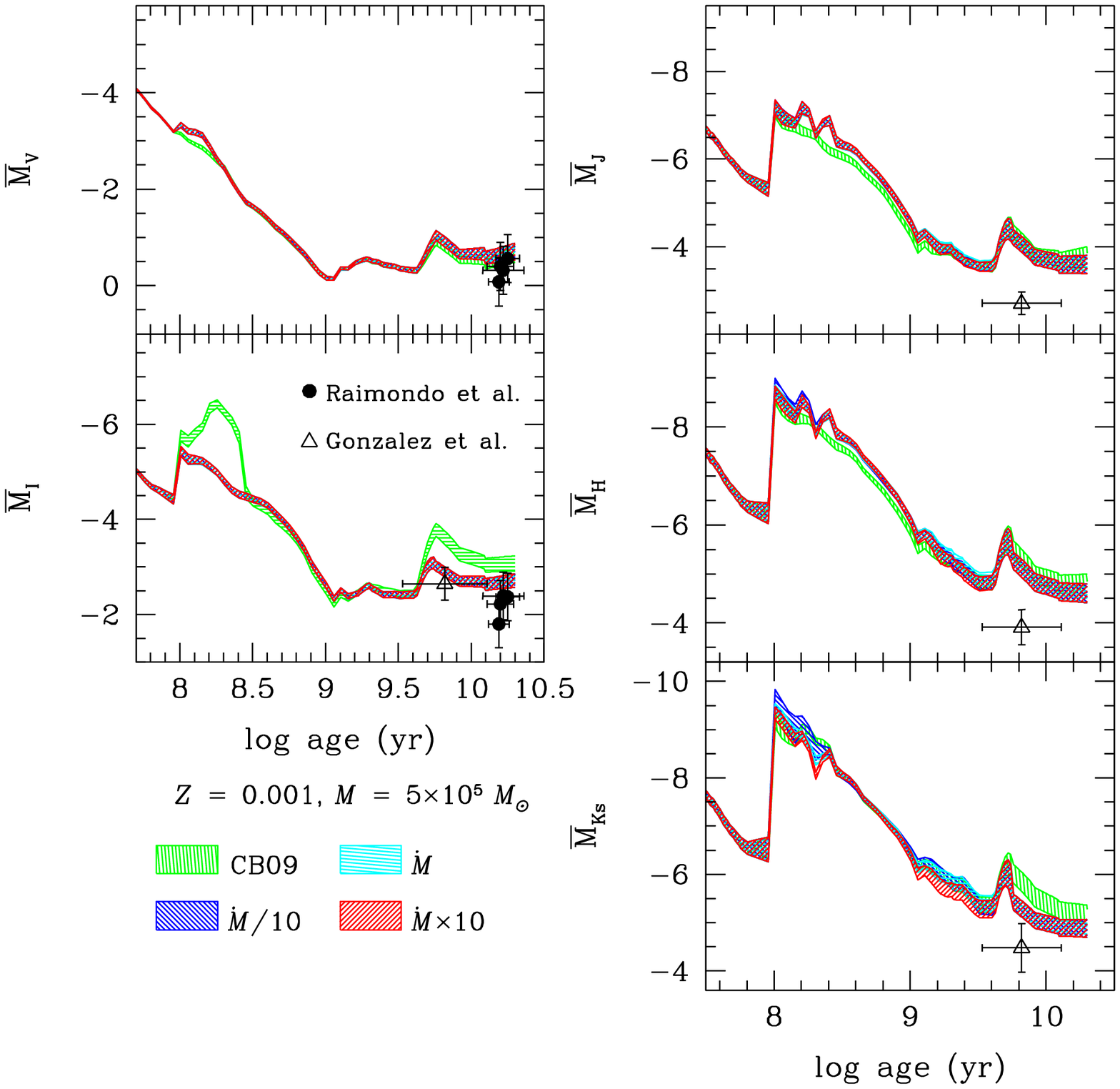}
\caption{SBFs of Magellanic star clusters IV. Observations of old MC
star clusters are compared to models with $Z$ = 0.001.  $\pm 1 \sigma$
stochastic uncertainties are shown for stellar populations with $5
\times 10^5~M_\odot$, coded as in Figure \ref{figZ52}.  According to
their reported ages, Raimondo's four oldest MC clusters belong in this
figure.  }
\label{figclust22}
\end{figure*}

Given the significantly greater depth of their {\sl HST} data, \citeauthor{raim05} 
were able to measure both the numerator and denominator of equation
\ref{theequation} using resolved stars. They determined absolute $V$ and $I$ fluctuation magnitudes 
with the same LMC distance modulus ${(m-M)_0 = 18.5}$.
The comparison of the described cluster data with the results of the models
presented in this work is shown in Figures \ref{figclust52} to \ref{figclust22}. 

Figures \ref{figclust52} and  \ref{figclust42}\footnote{
Note that the x-axes in figures \ref{figclust42}, \ref{figclust32}, and \ref{figclust22}
start only at log (age) =  7.4, in order to highlight the features of the models
at older ages.}  
present $V$, $I$, $J$, $H$, and $K_s$ SBF absolute magnitudes 
vs.\ log (age) of young 
and intermediate-age MC clusters.
Optical ($\barM_V$ and $\barM_I$) measurements of individual, 
rich clusters have been taken from \citet{raim05}.
These data are shown as solid circles.
For this work, we have recalculated near-IR fluctuation values for
the superclusters as described above, and calculated $\barM_I$, as well.
These data are shown as empty triangles.
As seen before, SBF measurements are not capable of discriminating
between different mass-loss rates.

In the near-IR, there is a good overall match between
models and data.
In the optical $V$ and $I$-bands, 
there is a tendency for the data
of single MC clusters first published by \citet{raim05} to fall below the model values.
Among these clusters, the problem is most acute 
for NGC 1805 (log $t = 7.00 \pm 0.20$), NGC 1818 (log $t = 7.40 \pm 0.30$), and
Hodge 14 (log $t = 9.30 \pm 0.10$). In the case of Hodge 14, severe 
field contamination forced \citeauthor{raim05} to analyse only a small area (i.e., 
mass) of the cluster; the resulting larger stochastic uncertainty 
is reflected by the SBF magnitudes error bars. For NGC 1805 and 
NGC 1818, they mention a possible 10\% incompleteness of 
the brightest 3 mag of cluster stars within 7$\farcs2$ from the centre, 
that in principle is accounted for by the shown error bars. 
A logical conclusion --also reached by 
\citet{raim05}-- is that the number of stars in these
individual globular clusters is not enough to adequately sample the brightest, rarest, 
TP-AGB stars. Additional systematic effects that might 
impact the data of both NGC 1805 and the pre-SWB supercluster of
\citeauthor{gonz04} have been already discussed in section 
\ref{sbfdat}.
  
Results for clusters with $Z \sim$ 0.002 are shown in Fig.~\ref{figclust32}.
Figure \ref{figclust22} compares SBF measurements of old MC clusters,
with $Z \sim$ 0.0007, to models with $Z$ = 0.001. The agreement
between models and data is roughly the same if models with $Z$ =
0.0004 are used instead. For the one ``supercluster" shown and 
contrariwise to what
we had seen so far, we find that the fit between models and data in
the optical is better than the match in the near-IR.
Since \citet{raim05} do not use our same mapping between SWB type and age (and 
thus metallicity),
we show their four oldest clusters in both figures \ref{figclust32} and 
\ref{figclust22}. Their quoted SWB type is VI, like our one supercluster 
displayed in Fig.~\ref{figclust32} as an empty triangle, whereas 
their listed ages are all over 10 Gyr old, i.e., 
older than our type VII supercluster,
and hence we plot them too in Fig.~\ref{figclust22}.

\section{Mid-infrared SBF measurements and future work.} \label{midirsbf}

\begin{figure*}
\includegraphics[width=1.\hsize]{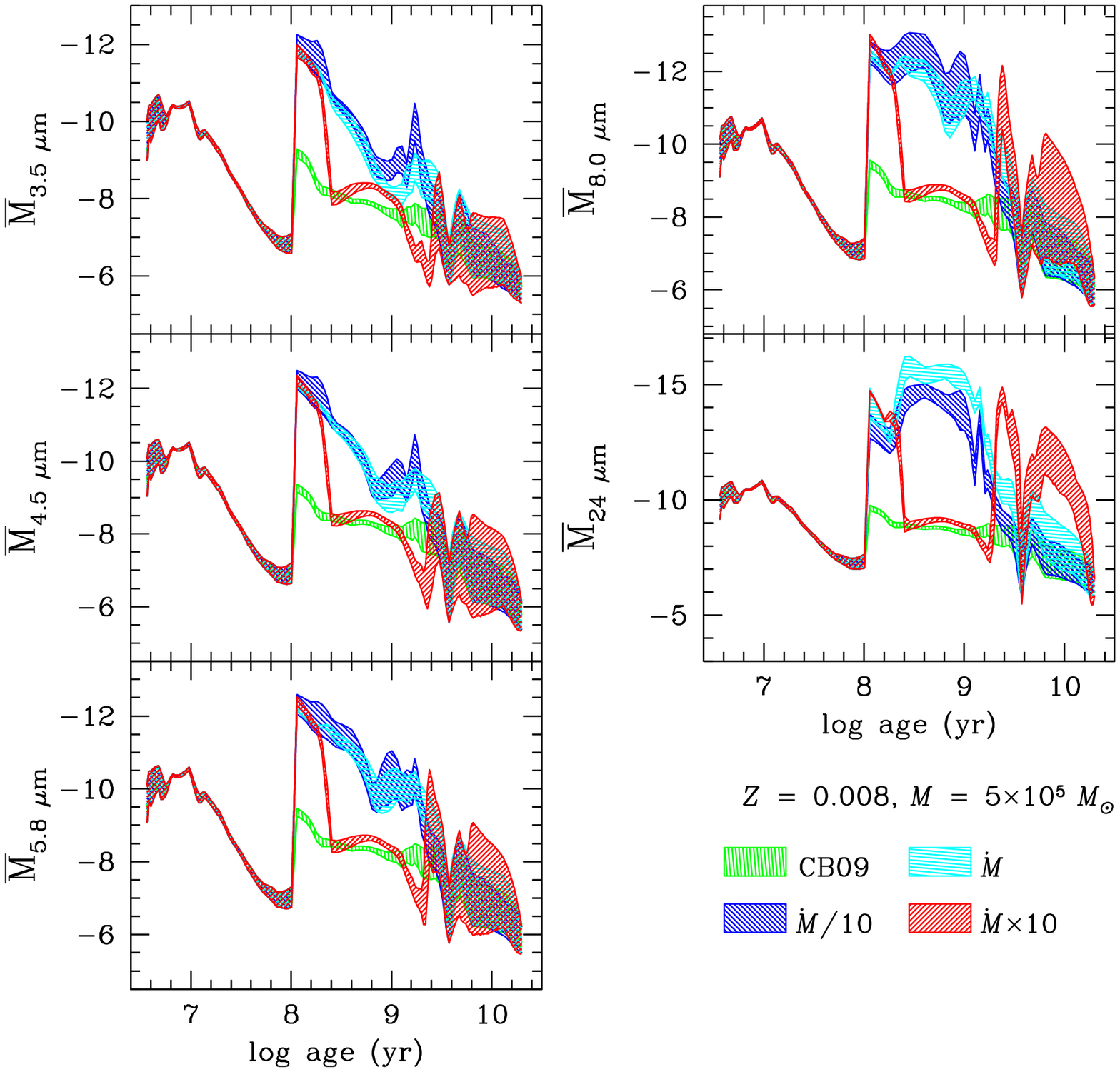}
\caption{ Mid-IR absolute fluctuation magnitudes vs.\ log (age) for
$Z$ = 0.008; models with different mass-loss rates.  Coloured regions,
coded as in Figure \ref{figZ52}, delimit expected $\pm 1 \sigma$
stochastic errors for stellar populations with $5 \times
10^5~M_\odot$.  }
\label{spitzer}
\end{figure*}

So, can SBF measurements at all, with their sensitivity to the brightest stars of a population,
provide some insight about the mass-loss parameters of unresolved stellar populations?
Figure \ref{spitzer} shows absolute fluctuation magnitudes vs. log (age), again for $Z$ = 0.008,   
but this time in the mid-IR bands observed by the Spitzer Space Telescope. 
According to this figure, in the mid-IR one could begin to distinguish 
intermediate age stellar populations with different mass-loss rates, to the
point that it might be  
worthwhile to start exploring the effects of other dust mixtures and 
even of the dust chemistry in the stellar envelopes on the integrated 
properties of stellar populations. In a forthcoming paper, we will compare
models in the mid-IR with Spitzer observations of stars in the Large Magellanic
Cloud. 

\section{Summary and conclusions.}

We have presented optical and IR broad-band colours and SBF magnitudes computed from 
single stellar population synthesis models, where the 
mass-loss rates in the CB09 evolutionary tracks have been used to produce 
spectra of TP-AGB stars, taking into account the radiative 
transfer in their dusty circumstellar envelopes. Star plus envelope SEDs
have been processed 
also for $\dot M$ one order of magnitude above and below the fiducial mass-loss rates; 
for mass-loss rates different from the original ones in the tracks, stellar configurations and lifetimes 
have been adjusted. 
Next, we have compared our models to optical and near-IR data of 
single AGB stars and Magellanic star clusters.

Even though mass-loss regulates lifetimes, luminosities, and effective temperatures
of stars in the AGB phase, and hence their numbers and colours, 
it turns out that  
broad-band optical and near-IR colours and SBF measurements of 
stellar populations cannot discern global variations in mass-loss rate.
Worse even than for single stars, the selection effects that preclude detection
of single stars away from the ``cliff" in a core mass vs.\ luminosity plot (stars either 
losing mass too slowly for the mass-loss to be observed or too fast for the star
to be detected, 
due both to lifetime and obscuration) make 
differences in mass-loss rate across whole populations completely unnoticeable.\footnote{
According to a very recent article \citep{raim09} that appeared while
the present work was being refereed,
an extinction $A_V$ between 0.01 and 1 mag 
is adequate to fit 
near-IR data of AGB stars. Raimondo 
concludes that dust-enshrouded AGB stars, with higher opacities, would rather 
leave their imprint at longer wavelengths. 
}

We predict that SBF measurements in the mid-IR could begin to pick out
intermediate age stellar populations where the stars lose mass with rates
away from the cliff line, and help determine if such deviations actually
correlate with metallicity.

\section*{Acknowledgments}
It is our pleasure to thank three anonymous referees for their
very thorough and helpful reports.
R.A.G.-L. recognizes the support of CONACyT, M\'exico, and  
DGAPA, UNAM. She also thanks M.\ Matsuura, J.\ Blakeslee, P.\ D'Alessio,
J.\ van Loon, K.-P.\ Schr\"oder, S.\ Kurtz, L.\  Piovan, and M.\ Cervi\~no.
Sundar Srinivasan and Margaret Meixner made their catalog of 
AGB candidates in the Magellanic Clouds available to us in
advance of publication.

We thank the whole DENIS Team, and especially its PI, N.\ Epchtein,
for making available de DENIS data. The DENIS project is supported,
in France by the Institut National des Sciences de l'Univers, the
Education Ministry and the Centre National de la Recherche Scientifique,
in Germany by the State of Baden W\"urtemberg, in Spain by the
DGICYT, in Italy by the Consiglio Nazionale delle Ricerche, in Austria
by the Fonds zur F\"orderung der wissenschaftlichen Forschung and
the Bundesministerium f\"ur Wissenschaft und Forschung.

This research has made use of the VizieR Service
and the SIMBAD database at
the Centre de Donn\'ees Astronomiques de Strasbourg, as well as NASA's
Astrophysics Data System Abstract Service.

\appendix

\section{Modelling the stellar plus envelope emission.} \label{modenv}

\subsection{Mass-loss rate and stellar parameters.} \label{dotmnstarpars}

We treat $\dot M$ as the independent parameter, and modify 
$L$, using Figure 2 in \citet{bowe91} \citep[reproduced as Figure 14 in][]{mari07}.
The former gives lines of constant $\dot M/M$
in the log~$M$ vs.\ log~$L$ plane, for dusty Miras with solar metallicity. 

Next, we find the modified stellar radius from 
the \citet{iben84} radius-luminosity-mass relation for 
evolving AGB stars, reproduced as equation 3 in \citet{bowe91},\footnote{
$R = 312(L/10^4)^{0.68}(M/1.175)^{-0.31S}(Z/0.001)^{0.088(l/H_p)^{-0.52}}$,
where $L$ and $M$ are in solar units, ($l/H_p$) is the ratio of mixing
length to pressure scale height, and $S = 0$ for $M \leq 1.175$ and  
$S = 1$ otherwise.
}
and calculate the effective
temperature using $L = R^2(T_{\rm eff}/5770)^4$. 

This is the inverse of the procedure used by \citet{mari07} to determine $\dot M$
for an oxygen-rich AGB star (from mass and $Z$ they calculate period, next radius, then luminosity,
and ultimately mass-loss rate).

Once the new $L$, $R$, and $T_{\rm eff}$ are known, and assuming the energy available to a 
star of type $i$ during an 
evolutionary phase is constant \citep[i.e., the fuel consumption theorem,][]{renz86}, the length $t_2$ of the extrapolated
mass-loss phase can be derived 
from the relation $L_1 t_1$ = $L_2 t_2$, where
subscript 1 denotes original parameters and subscript 2 represents the 
extrapolated ones. If, furthermore, we assume that the weight $w_i$ of a stellar type $i$
(i.e., the number of stars of initial mass $m$ going through an 
evolutionary phase $ph$, per unit mass of the population) is proportional to the time 
spent by such stars in the phase, then  
$w_{i,2}/w_{i,1} = t2/t1 = L1/L2$. $w_{i,1} \equiv 1$, by definition, and $w_{i,2}$ is the same for all stars with the same
metallicity, once the extrapolation factor is fixed, with the following exceptions.
There are occasions when the product of the mass-loss rate and the  
phase length derived with the above procedure is greater than the 
stellar core mass at the beginning of the stage, $M_{\rm evst}$. In these cases, $w_{i,2} = M_{\rm evst}/\dot M/t_2$, and 
the weight of the subsequent stage will be zero (i.e., the star will not reach the
following stage). In other instances, the dust-to-mass ratio $\Psi$ diverges (see eq.~\ref{psieq2} in Section~\ref{dtgsec} below),  
consequently the shell opacity also diverges (see eq.~\ref{taueq}, Section~\ref{dustenv}), and hence we assign
$w_{i,2} = 0$ to such stars in view of their very heavy obscuration.  
In general, the mass lost in any given stage with modified mass-loss rate will be different from 
the mass lost in the original track, so an adjustment to the parameters of the subsequent stage will
be needed. Logically, we first rectify the stellar mass $M_{\rm evst}$ and next the luminosity $L_1$,
moving along a line of constant $\dot M/M$ in the aforementioned Figure 2 of \citet{bowe91}.
From then on, we repeat the procedure outlined here: we first modify $\dot M$, 
and then derive $L_2$, $R_2$, $T_{eff,2}$, $t_2$, $w_{i,2}$.

Fundamental mode (FM) pulsation periods $P_0$ also change with the stellar configuration.
To adjust these, we use eq.~12 in \citet{mari07}:

\begin{eqnarray}
{\rm log}\ (P_0/{\rm days}) = -\ 2.07\ +\ 1.94\ {\rm log}\ (R/R_\odot) \nonumber \\
-\ 0.9\ {\rm log}\ (M/M_\odot)\ ({\rm if}\ M\ <\ 1.5 M_\odot) \nonumber \\
                            = -\ 2.59\ +\ 2.2\ {\rm log}\ (R/R_\odot) \nonumber \\
-\ 0.83\ {\rm log}\ (M/M_\odot)\ -\ 0.08\ {\rm log}\ (Z/10^{-3}) \nonumber \\
+\ 0.25(Y - 0.3)\ ({\rm if}\ M\ >\ 2.5 M_\odot). 
\label{eqperiod}
\end{eqnarray}

\noindent
$Z$ and $Y$ are, respectively, metallicity and helium content by mass; for $M$ between 1.5 and 2.5 $M_\odot$, 
log $P_0$ is interpolated linearly, with log $M$ as the independent variable. 
In the case of C-rich stars in the superwind phase, another parameter that changes with $\dot M$ is the C/O ratio.
For this, we use eq.~23 in \citet{mari07}:\footnote{
This expression is similar to that proposed by
\citet{wach02} for winds driven by radiation pressure on dust grains, except that it contains an explicit
dependence on the C/O ratio. Eq.~23 of \citet{mari07} has an error:
the minus sign in front of log $P_0$ should be a plus, as it appears here and in \citet{wach02}.
}

\begin{eqnarray}
{\rm log}\ [\dot M/(M_\odot {\rm yr}^{-1})] = -\ 4.529 \nonumber \\ 
 -\ 6.849\ {\rm log}\ (T_{\rm eff}/2600\ {\rm K})\ +\ 1.527\ {\rm log}\ (L/10^4 L_\odot) \nonumber \\ 
 -\ 1.997\ {\rm log}\ (M/M_\odot)\ +\ 0.995\ {\rm log}\ (P_0/650\ {\rm days}) \nonumber \\ 
 +\ 0.672\ {\rm log}\ \left(\frac{C/O}{1.5}\right). 
\label{eqcorat}
\end{eqnarray}

\noindent Eq.~\ref{eqcorat} is valid for $\dot M \geq 10^{-6}\ M_\odot$ yr$^{-1}$.

\subsection{Parameters of the dusty envelopes.} \label{dustenv}

The effect of dust on the stellar SEDs has to be included, both for stars in the original tracks 
and for those whose mass-loss rates and stellar parameters have been modified.
To this end, we follow the procedure outlined by \citet{piov03} and \citet{mari08} based on,
respectively, \citet{ivez97} and \citet{elit01}.
One very important input parameter to produce the SED through 
radiative transfer calculations 
is the optical depth, $\tau$, of the envelope.

For a star of luminosity $L$ and effective temperature $T_{\rm eff}$, losing mass at a rate $\dot M$, with dust expansion velocity $v_{\rm exp}$ and 
dust-to-gas ratio $\Psi$, the optical depth at wavelength $\lambda$ is approximately
\citep[see eq.~12 of][]{piov03}:

\begin{equation}
\tau_\lambda \simeq \frac{\Psi \dot M \kappa_\lambda}{4 \pi v_{\rm exp}} \frac{1}{r_{\rm in}}, 
\label{taueq}
\end{equation}

\noindent where $\kappa_\lambda$ is the extinction coefficient per unit mass and $r_{\rm in}$ is the dusty
shell internal radius. 

If the shell is optically thick to radiation of wavelength $\lambda$, we can write

\begin{equation}
L = 4 \pi R^2 \sigma T^4_{\rm eff} = 4 \pi r^2_{\rm in} \sigma T^4_{\rm d},
\end{equation}

\noindent and

\begin{equation}
r_{\rm in} = \left( \frac{L_\odot}{4 \pi \sigma T^4_{\rm d}} \right)^{1/2} \left(\frac{L}{L_\odot} \right)^{1/2},
\label{eqtaupiov}
\end{equation}

\noindent where $R$ is the stellar radius, $\sigma$ is the Stefan-Boltzmann 
constant, and $T_{\rm d}$ is the dust condensation temperature at $r_{\rm in}$.
Adopting $T_{\rm d} =$ 1000 K when C/O $<$ 0.97 and  $T_{\rm d} =$ 1500 K when C/O $\geq$ 0.97, 

\begin{eqnarray}
r_{\rm in} = 2.37 \times 10^{12} \left( \frac{L}{L_\odot} \right )^{1/2} {\rm cm}\ ({\rm for\ O-rich\ stars})  \nonumber \\ 
r_{\rm in} = 1.05 \times 10^{12} \left( \frac{L}{L_\odot} \right )^{1/2} {\rm cm}\ ({\rm for\ carbon\ stars});
\end{eqnarray}

\noindent
in any case \citep[see eq.~4 of][]{mari08}:

\begin{equation}
\tau \propto \dot M \Psi v^{-1}_{\rm exp} L^{-0.5}.
\end{equation}

\subsubsection{Wind expansion velocity.}

Based on the solution to the dust-wind problem by \citet{elit01},  \citet{mari08} express 
the wind expansion velocity $v_{\rm exp}$ as a function of $\dot M$,  $L$, $\Psi$, and 
other dust parameters as follows:

\begin{equation}
v_{\rm exp} = \left (A \dot M_{-6} \right)^{1/3} \left(1 + B \frac{\dot M^{4/3}_{-6}}{L_4}\right)^{-1/2}\ {\rm km\ s}^{-1},
\label{eqvexp}
\end{equation}

\noindent
where $\dot M_{-6}$ is the mass-loss rate in units of 10$^{-6} M_\odot$ yr$^{-1}$ and $L_4$ is the stellar
luminosity in units of 10$^4 L_\odot$.

The parameters $A$ and $B$ are defined as:

\begin{eqnarray}
A = 3.08\times 10^5\ T^4_{\rm d3}\ Q_\star \sigma^2_{22}\ \chi^{-1}_0 \\
B = \left[2.28 \frac{Q^{1/2}_\star \chi^{1/4}_0}{Q^{3/4}_V \sigma^{1/2}_{22} T_{\rm d3}} \right]^{-4/3}.
\end{eqnarray} 

\noindent
$T_{\rm d3}$ is the dust condensation temperature in units of 10$^3$ K; $Q_\star$ is the Planck average\footnote{
The Planck mean of a function  $Q(\lambda)$ is given by \citep{blan83}:

\begin{equation}
<Q(T)> = \frac{\int\limits^\infty_0 B(\lambda,T)Q(\lambda) {\rm d}\lambda}{\int\limits^\infty_0 B(\lambda,T) {\rm d}\lambda},
\end{equation}
where $B(\lambda,T)$ is the Planck function at a temperature $T$.
}
of the efficiency coefficient for radiation pressure,
evaluated at $T_{\rm eff}$,
and $Q_V$ is the 
efficiency coefficient for absorption at the visual range. $\chi_0$ is defined as:

\begin{equation}
\chi_0 = \frac{Q_P(T_{\rm eff})}{ Q_P(T_{\rm d})}, 
\end{equation} 

\noindent with $Q_P(T)$ the Planck average of the absorption efficiency at temperature $T$, 
whereas $\sigma_{22}$ is $\sigma_{\rm gas}$, the dust cross-section per gas particle, in units of 10$^{-22}$ cm$^2$, and 

\begin{equation}
\sigma_{\rm gas} = \pi \left(\frac{a}{2}\right)^2 \frac{n_{\rm dust}}{n_{\rm gas}}\ {\rm cm}^2,
\label{siggaseq}
\end{equation}

\noindent
with $a$ the mean size of the dust grains in cm (assumed to be the same 
for all grains), and $n_{\rm dust}$ and $n_{\rm gas}$
the number densities in cm$^{-3}$ of, respectively, the dust and gas particles. 

$\sigma_{\rm gas}$ can be written as a function of $\Psi$, the dust-to-gas ratio, as follows:

\begin{equation}
\Psi = \frac{\rho_{\rm dust}}{\rho_{\rm gas}} = \frac{\frac{4}{3} \pi \left(\frac{a}{2}\right)^3 \rho_{\rm grain} n_{\rm dust}}{A_{\rm gas} m_H n_{\rm gas}}, 
\label{psieq}
\end{equation}

\noindent
where $\rho_{\rm dust}$ and $\rho_{\rm gas}$ are the density of matter in the form
of dust and gas, respectively, in g cm$^{-3}$. 
$A_{\rm gas} \simeq 4/(4 X_{\rm H} + X_{\rm He}) $ is the mean molecular weight of the gas 
in units of the H atom mass, $m_{\rm H}\ =\ 1.674 \times 10^{-24}$ g, assuming that all the
gas is composed by H and He, with respective mass fraction abundances $X_{\rm H}$ and $X_{\rm He}$. 
Substituting eq.~\ref{psieq} into eq.~\ref{siggaseq},

\begin{equation}
\sigma_{\rm gas} = \frac{3}{2} \frac{A_{\rm gas} m_{\rm H}}{a\ \rho_{\rm grain}} \Psi.
\end{equation}

The values of $Q_V$ for silicates, silicon carbide, and amorphous carbon (see below, Subsection 
\ref{dtgsec} and Section \ref{impdusty}) are taken from Table 3 in 
\citet{mari08}; the values of $Q_\star(T_{\rm eff})$, $Q_P(T_{\rm eff})$, and $Q_P(T_{\rm d})$
(each one for all the same materials) are interpolated from the quantities tabulated there, 
taking temperature as the independent variable.\footnote{As seen in Section~\ref{dustenv},
$T_{\rm d} =$ 1000 K for C/O $<$ 0.97, and  $T_{\rm d} =$ 1500 K for C/O $\geq$ 0.97.} 

Clearly,  eq.~\ref{eqvexp} is valid as long as there is enough dust condensation to drive an 
outflow. This condition sets the minimum mass-loss rate \citep{elit01}: 

\begin{equation}
\dot M_{\rm min} = 3 \times 10^9 \frac{M^2}{Q_\star \sigma^2_{22} L_4 T^{1/2}_{k3}}\ M_\odot\ {\rm yr}^{-1}.  
\end{equation}

\noindent
$T_{\rm k3}$ is the kinetic temperature (in units of 10$^3$ K) at the inner boundary of
the dust shell, which is assumed to be equal to the dust condensation temperature, 
$T_{\rm d3}$. At smaller values of $\dot M$, dust cannot drive a wind but 
may still form in the circumstellar envelope. Following \citet{mari08},  
we handle this situation by setting $v_{\rm vexp} = v_{\rm exp}(\dot M_{\rm min})$, 
while using the actual $\dot M$ to 
calculate the opacity of the envelope.

\subsubsection{Dust-to-gas ratio.}  \label{dtgsec}

If dust and gas share the same outflow velocity, the dust-to-gas ratio can also
be written as the quotient between mass-loss in the form of dust and 
mass-loss in the form of gas:

\begin{equation}
\Psi = \frac{\dot M_{\rm dust}}{\dot M_{\rm gas}} = \frac {\dot M_{\rm dust}}{\dot M - \dot M_{\rm dust}}.
\label{psieq2}
\end{equation}

We calculate $\dot M_{\rm dust}$ as described by \citet{mari08}, who follow the formalism by 
\citet{ferrarotti03,ferrarotti06}. Summarizing, 

\begin{equation}
\dot M_{\rm dust} = \sum_i \dot M_{{\rm dust},i}, 
\end{equation}

\noindent 
with the summation over several dust species, written as 

\begin{equation}
\dot M_{{\rm dust},i} = \dot M X_{\rm seed} \left(\frac{A_{{\rm dust},i}}{A_{\rm seed}}\right) f_{{\rm dust},i}.
\end{equation} 

\noindent
Here, $A_{{\rm dust},i}$ is the mean molecular weight of the dust
species, $X_{\rm seed}$ is the total mass fraction of the seed element
in the circumstellar shell, and $A_{\rm seed}$ is its atomic
weight. The seed element is the least abundant one among those needed
to form the considered dust species, and hence limits its supply.
Finally, $f_{{\rm dust},i}$ is the fraction of the seed element
condensed into dust grains, that for each dust species is calculated
as a function of $\dot M$ and C/O ratio using the analytic fits in
\citet{ferrarotti03}, as shown below.

\subparagraph{Stars with C/O $<$ 0.97.} \label{miras}

In these conditions, there are enough oxygen atoms to form
silicate-type dust, and:

\begin{equation}
\frac{dM_{\rm dust}}{dt} = \frac{dM_{\rm sil}}{dt} = \dot M X_{\rm Si}
\frac{A_{\rm sil}}{A_{\rm Si}} f_{\rm sil},
\end{equation} 

\noindent
where $A_{\rm Si}$ is the atomic weight of silicon, $A_{\rm sil}$ is
the effective molecular weight of the silicate dust, and $f_{\rm sil}$
is the condensation degree, also of the silicate dust.  Given the dust
mixture we use here for O-rich stars (see below, Section~\ref{impdusty}),
the condensation degree includes the contributions from amorphous and
crystalline silicates:

\begin{equation}
f_{\rm sil} = f_{\rm warm} + f_{\rm cold} + f_{\rm ens} + f_{\rm fors}, 
\end{equation}

\noindent 
where subscripts stand for warm silicate, cold silicate, enstatite,
and forsterite dust.  On the other hand, the effective molecular
weight of the silicate dust mixture is:

\begin{equation}
A_{\rm sil} = (f_{\rm warm} A_{\rm warm} + f_{\rm cold} A_{\rm cold} + f_{\rm ens} A_{\rm ens} + f_{\rm fors} A_{\rm fors})/f_{\rm sil}.
\end{equation}

The degree of condensation of silicate type dust is found from the
analytic fit in \citet{ferrarotti03}, used also by \citet{mari08}:

\begin{equation}
f_{\rm sil} = 0.8 \frac{\dot M_{-6}}{\dot M_{-6} + 5} \sqrt\frac{Y_{\rm C,1} - Y_{\rm C}}{Y_{\rm C,1}}.
\end{equation}

\noindent 
Here, $Y = X/A$ is abundance in molecules g$^{-1}$, and $Y_{\rm C,1} =
Y_{\rm O} - 2Y_{\rm Si}$ \citep{ferrarotti06}.  Once $f_{\rm sil}$ is
determined for each star, the relative degrees of condensation of
crystalline and amorphous silicates are set according to the optical
depth of the envelope, as described below (Section~\ref{impdusty}).\footnote{ For the cold and
warm amorphous silicates we use the following formulae, respectively
\citep{dors95}: Mg$_{0.8}$Fe$_{1.2}$SiO$_4$;
Mg$_{0.4}$Fe$_{0.6}$SiO$_3$.  }

\subparagraph{Stars with C/O $\geq$ 0.97.}  \label{carbonstars}

These are C-rich stars, since the 
dust mixture in their envelopes 
is dominated by carbon. As stated below in Section~\ref{impdusty}, we consider two dust constituents, 
SiC and amorphous carbon (AMC), such that: 

\begin{equation}
\frac{dM_{\rm dust}}{dt} = \frac{dM_{\rm SiC}}{dt} + \frac {dM_{\rm AMC}}{dt}, 
\end{equation}

\noindent
where the first addend corresponds to the silicon carbide dust and the second,
to the amorphous carbon dust. 
Furthermore,

\begin{eqnarray}
\frac{dM_{\rm AMC}}{dt} = \dot M X_{\rm C} f_{\rm AMC}.
\end{eqnarray}

\noindent
$X_{\rm C}$ is the carbon abundances by mass; we take the current atmospheric value, solar 
scaled, or $X_{\rm C} = ({\rm C/O}) X_{{\rm O},\odot} (Z/Z_\odot)$, 
where $X_{{\rm O},\odot}$ is the solar mass fraction abundance of oxygen
taken from \citet{grev98}.\footnote{For consistency with
\citet{mari07}, and with the different works by Ferrarotti and collaborators.} 

Finally, the degree of condensation of carbon is found from \citet{ferrarotti03}:\footnote{Ferrarotti writes
(C/O - 1), but we change it to (C/O - 0.97), to avoid the possibility of negative degrees of condensation.}

\begin{equation}
f_{\rm AMC} = 0.5 \frac{\dot M_{-6}}{\dot M_{-6} + 5} \left( \frac{\rm C}{\rm O} - 0.97 \right).   
\end{equation} 

\noindent
Once $f_{\rm AMC}$ is found, the relative degree of condensation of SiC dust is set according to 
the opacity of the envelope, as described below (Section~\ref{impdusty}). 
 
\subsubsection{Extinction coefficient.}

In order to determine $\kappa_\lambda$, we follow the procedure by \citet{piov03}. 
The mass extinction coefficient can be written as:

\begin{equation}
\kappa_\lambda = \frac{\sum_i n_{{\rm dust},i} \sigma_{{\rm dust},i}} {\rho_{\rm dust}},
\end{equation}  

\noindent
with the summation over all types of grains in the dust mixture, and where 
the $i$th type of grain has number density 
$n_{{\rm dust},i} $ cm$^{-3}$,
and cross-section for radiation-dust interactions $\sigma_{{\rm dust},i}$ cm$^2$; 
$\rho_{\rm dust}$ (g cm$^3$) is, again, the density of matter in the form of dust, and  
$\sigma_{{\rm dust},i} (a) = \pi a^2 Q_{\rm ext} (i)$, where $Q_{\rm ext} (i)$ are
the extinction coefficients. If $m_{{\rm dust},i}$ is the mass of the 
$i$th type of dust grains, and we introduce the mass abundance $\chi_i = n_{{\rm dust},i} m_{{\rm dust},i} / \rho_{\rm dust}$: 

\begin{equation}
\kappa_\lambda = \sum_i \chi_i \frac{\sigma_{{\rm dust},i}}{m_{{\rm dust},i}}.
\end{equation}

Going back to eq.~\ref{taueq}, one can see that dust optical depth is a function of $\kappa_\lambda$, 
but then $\kappa_\lambda$ is a function of the optical depth, because the dust composition
is itself a function of $\tau_\lambda$. Consequently, an iterative procedure is needed to
determine the dust optical depth.

\subsection{Implementation of the envelopes with DUSTY.}\label{impdusty}

The DUSTY code includes the dependence of the wind speed on 
stellar luminosity and metallicity (through the gas-to-dust ratio of
the envelope), and the drift speed between the dust and the gas.
The code outputs spectral {\it shapes}, that can then be scaled to
energy flux by multiplying by the stellar luminosity, and dividing 
by 4$\pi d^2$. 
DUSTY offers a broad choice of input parameters, specifically in 
regard to the dust chemical composition and grain size distribution.

\begin{figure*}
\includegraphics[width=0.70\hsize,angle=-90,clip=]{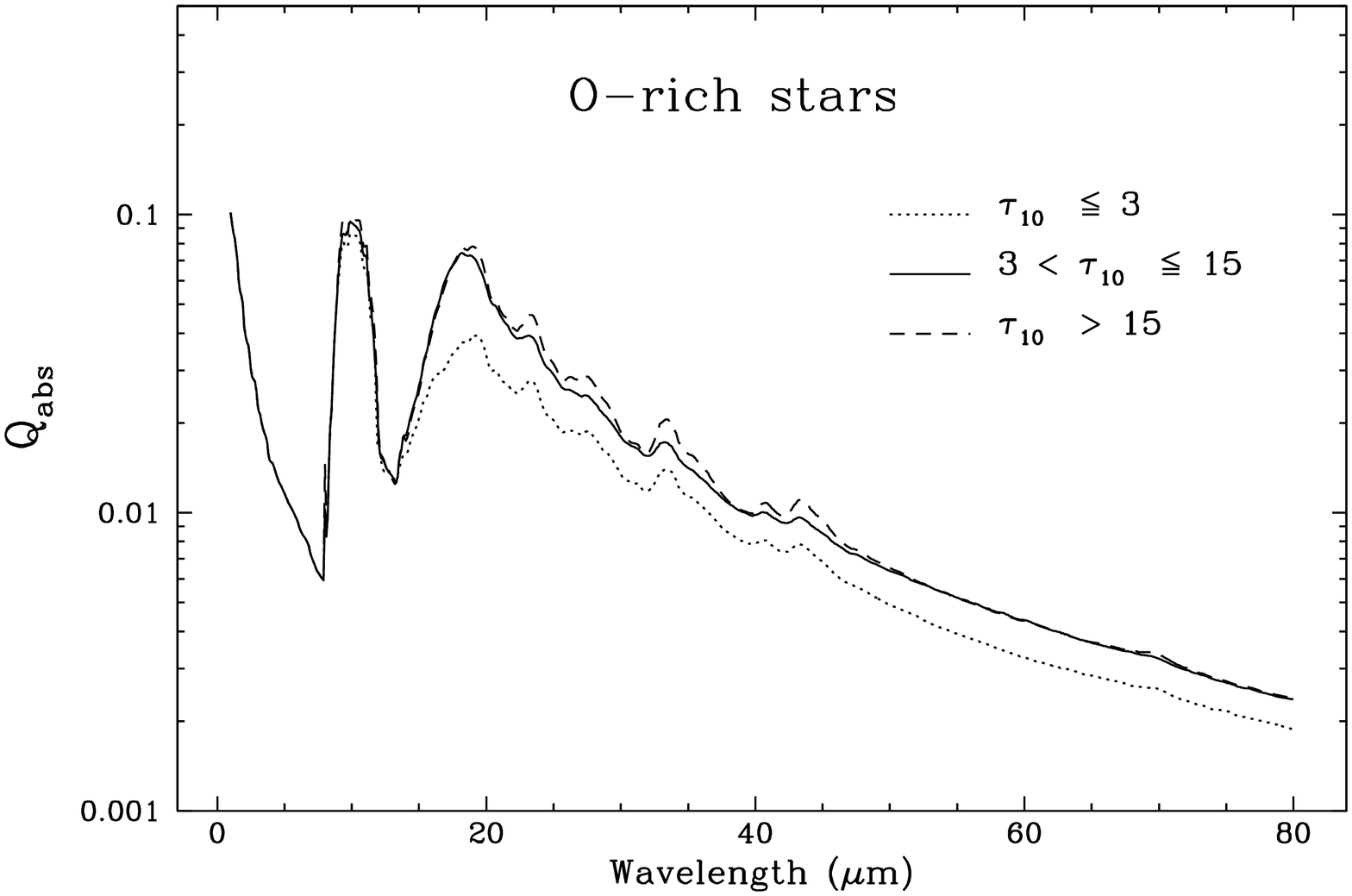}
\caption{Absorption efficiency functions of different
mixtures of siliceous dust grains, for  
envelopes of O-rich stars with various optical depths. 
{\it Dotted-line:} $\tau_{10} \leq$ 3; {\it solid-line:} 
3 $ < \tau_{10} \leq$ 15; {\it dashed-line:} 
$\tau_{10} > $ 15 (see text). 
}
\label{qiuo}
\end{figure*}

We adopt here the dust mixtures 
introduced by \citet{suh99,suh00,suh02},
with the aim of fitting observations of individual stellar spectra,
and later used by \citet{piov03} in SSP models. 
Suh has calculated optical constants for siliceous and carbonaceous
compounds that are consistent with the 
Kramers-Kronig dispersion relations and, at the same time, yield 
models of dusty envelopes that fit observed properties of AGB
stars.  
For O-rich stars, \citet{suh02} proposes combinations of 
both amorphous and crystalline silicate grains (enstatite, MgSiO$_3$, and 
forsterite, Mg$_2$SiO$_4$), in proportions that depend 
on the optical depth of the dusty shell: $(a)$ for stars with the lowest mass-loss
rates and thin shells, where the 10 $\mu$m silicate feature is
observed in emission ($\tau_{10} \leq$ 3), 
a blend of 90\% warm amorphous silicate, 5\% enstatite, and 5\% forsterite grains; 
$(b)$ for stars with low mass-loss rates and moderately optically thick 
shells (3 $ < \tau_{10} \leq$ 15), 
90\% cold amorphous silicate, 5\% enstatite, and 5\% forsterite grains;
$(c)$ for stars with high mass-loss rates and optically thick shells
($\tau_{10} > $ 15), 80\% cold amorphous silicate, 10\% enstatite, 
and 10\% forsterite. Dust opacity functions for these 3 cases 
are shown in Fig.\ \ref{qiuo}. 

\begin{figure*}
\includegraphics[width=0.70\hsize,angle=-90,clip=]{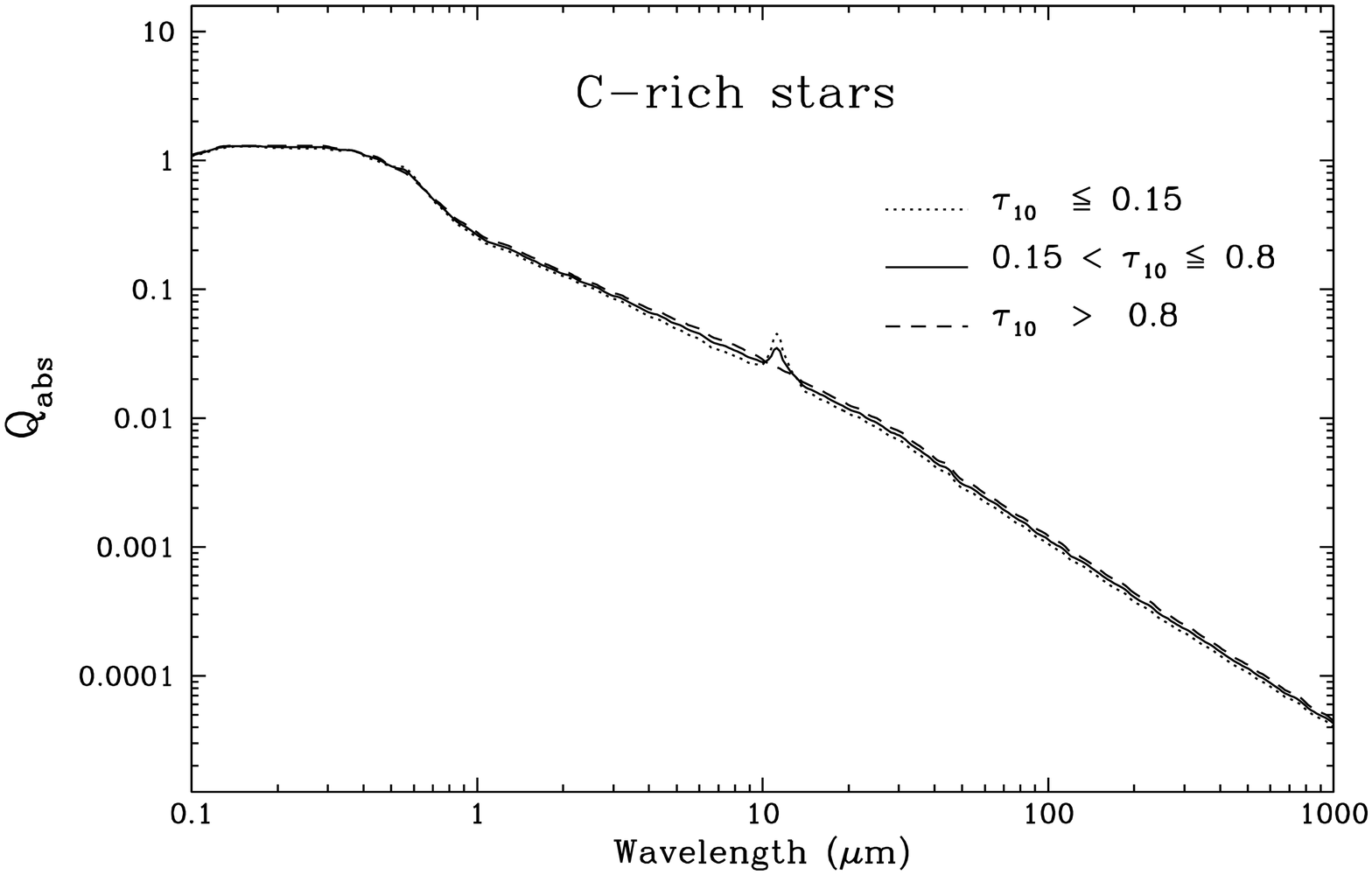}
\caption{Absorption efficiency functions of different 
mixtures of carbonaceous dust grains, for 
envelopes of C-rich stars with various optical depths. 
{\it Dotted-line:} $\tau_{10} \leq$ 0.15; {\it solid-line:} 
0.15 $ < \tau_{10} \leq$ 0.8; {\it dashed-line:} 
$\tau_{10} > $ 0.8 (see text).
}
\label{qiuc}
\end{figure*}

In the case of C-rich stars, \citet{suh00}
prescribes mixtures of amorphous carbon (AMC) and silicon carbide (SiC), 
again in ratios that depend on the optical depth of the dusty 
shell. $(a)$ For C-rich stars with optically thin shells ($\tau_{10} \leq$ 0.15),
20\% of SiC grains are needed to fit the strong 11 $\mu$m feature in 
observed spectra; $(b)$ 10\% of SiC grains are 
required for shells with intermediate optical depths 
(0.15 $< \tau_{10} \leq$ 0.8),
and $(c)$ no SiC grains are necessitated at larger optical depths, for which 
the 11 $\mu$m  feature is either weak or absent. Dust opacity 
functions for the 3 cases encountered for C-rich stars 
are displayed in Fig.\ \ref{qiuc}.
 
Optical constants for the different grains are taken from 
\citet[][ amorphous silicates]{suh99}, \citet[][ amorphous carbon]{suh00}, 
\citet[][ enstatite]{jaeg98}, and \citet[][ forsterite --actually, 
olivine: Mg$_{1.9}$Fe$_{0.1}$SiO$_4$]{fabi01}.
The data for $\alpha$-SiC come from \citet{pego88}, and are included 
in DUSTY's built-in library of optical constants. 
 
We take a uniform dust grain size of 0.1 $\mu$m, following 
\citet{piov03}. 
In general, for stars with $\dot M \neq$ 0, we use DUSTY's numerical solution for radiatively driven
winds, extending to a distance 10$^4$ times the inner radius. 
For $\tau \geq$ 80, however, and due to numerical 
difficulties of the program, we assume a shell density distribution 
that falls off as $r^{-2}$. 
As for the incident radiation (that DUSTY surmises comes from 
a point source at the centre of the density distribution), 
we use the same stellar SEDs as the BC03 and CB09 standard models
(adapted according to stellar parameters for modified mass-loss-rates),
and for the dust temperature on the envelope inner boundary we input 
1000 K for O-rich stars, and 1500 K for carbon stars
\citep{mari08}.\footnote{
For the most massive stars ($M \ga 4 M_\odot$), both
  C-rich and O-rich, occasionally there are extremely short ($10 - 100$
  yr long) phases with no mass-loss that follow mass-losing stages
  with a duration of $\sim 10^5$ yr.  We worried that standard (i.e.,
  Schultheis et al. or Westera et al., as appropriate) spectra would
  not provide a realistic modelling of such phases, and performed the
  experiment of assigning to them, instead, the spectra of the
  immediately preceding stage. The results of both procedures
  (standard spectra and preceding stage spectra) were
  indistinguishable for all practical purposes.}  For consistency with
\citet{suh99,suh00,suh02} and \citet{piov03}, we evaluate
$\kappa_\lambda$ and $\tau_\lambda$ at 10$\mu$m, and naturally all our
calculations with DUSTY take the same $\lambda$ as their reference
wavelength.


\label{lastpage}

\end{document}